\documentclass[twocolumn]{aastex62}
\usepackage{xcolor}
\usepackage{enumitem}

\usepackage{lipsum}
\usepackage{gensymb} 
\usepackage{mathtools}
\usepackage{placeins}
\usepackage{multirow}
\usepackage{float}
\usepackage{amsmath}
\usepackage{bm}
\usepackage{physics}
\usepackage{silence}
\usepackage{multirow, booktabs}
\newcommand{\analysisConfiguration}{configuration_MCMC_line_fit_function_extended_vol_PAH_bC_3_final}
\newcommand{\firstHypothesisAnalysisConfiguration}{configuration_fixed_volume_prior_fixed_AI_per_NH_final_1}
\newcommand{\optimizerGrid}{optimizer_grid_optimizer_combos_fixed_AI_1_3_bC_3}
\shorttitle{Dust extinction-emission correlation.}
\shortauthors{Zelko and Finkbeiner}

\begin{document}

\title{Implications of Grain Size Distribution and Composition for the Correlation between Dust Extinction and Emissivity}

\author{Ioana A. Zelko}
\affiliation{Harvard-Smithsonian Center for Astrophysics \\
60 Garden Street\\
Cambridge, MA 02138, USA}
\author{Douglas P. Finkbeiner} 
\affiliation{Harvard-Smithsonian Center for Astrophysics \\
60 Garden Street\\
Cambridge, MA 02138, USA}
\affiliation{Department of Physics, Harvard University \\
17 Oxford Street\\
Cambridge, MA 02138, USA}

\begin{abstract}
We study the effect of variations in dust size distribution and composition on the correlation between the spectral shape of extinction (parameterized by $R_{\textnormal{V}}$) and far-infrared dust emissivity (parameterized by the power-law index $\beta$). Starting from the size distribution models proposed by \cite{Weingartner2001a},  using the dust absorption and emission properties derived by \cite{Laor1993} for carbonaceous and silicate grains, and by \cite{Li2001} for polycyclic aromatic hydrocarbon grains, we calculate the extinction and compare it with the reddening vector derived by \cite{Schlafly2016}. An optimizer and an Markov chain Monte Carlo method are used to explore the space of available parameters for the size distributions.  
We find that larger grains are correlated with high $R_{\textnormal{V}}$. However, this trend is not enough to explain the emission-extinction correlation observed by \cite{Schlafly2016}. For the $R_{\textnormal{V}}-\beta$ correlation to arise, we need to impose explicit priors for the carbonaceous and silicate volume priors as functions of $R_{\textnormal{V}}$. The results show that a composition with higher ratio of carbonaceous to silicate grains leads to higher $R_{\textnormal{V}}$ and lower $\beta$. A relation between $E(\textnormal{B}-\textnormal{V})/\tau_{353}$ and $R_{\textnormal{V}}$ is apparent, with possible consequences for the recalibration of emission-based dust maps as a function of $R_{\textnormal{V}}$.

\end{abstract}

\keywords{interstellar medium, interstellar dust, interstellar dust extinction, interstellar dust processes}
\section{Introduction}\label{sec:intro}

\begin{figure*}[t]
	\centering
	\includegraphics[scale=1]{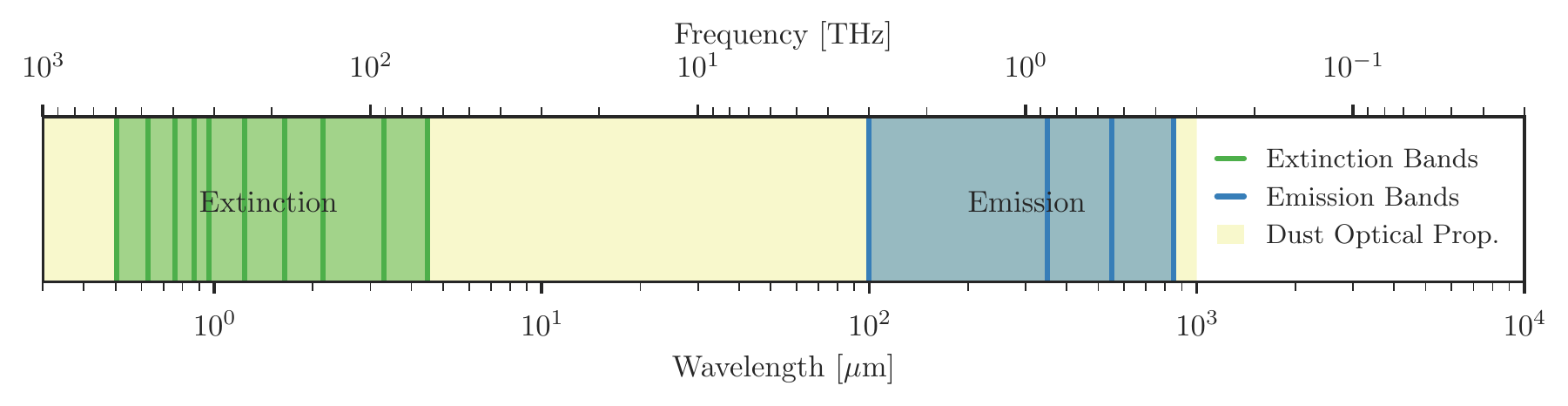}
	\caption{The extinction is evaluated at 10 bands of the reddening vector from S16. The emission is evaluated at the four bands used by the Planck satellite team:  353, 545, and 857GHz from Planck, and 3000GHz ($100 \mu$m) from IRAS. The extinction and emission wavelength ranges are far apart,  which raises the question of what drives the $R_{\textnormal{V}}-\beta$ relation. \label{fig:range}}
\end{figure*}

Dust is an important component of the interstellar medium, forming structures in the space between the stars in our galaxy. Dust is formed through the death process of stars, through supernovas or stellar winds. It is composed of elements that have formed in the stars, and it plays an important role in the further formation of complex molecules \citep{Draine2011}.

Dust scatters and absorbs ultraviolet and visible light coming from the interstellar radiation field (ISRF) around it. The scattering and absorption together cause extinction of the ultraviolet and visible light. Past studies have aimed to characterize the wavelength dependencies of the extinction. Their work \citep{Savage1979, Fitzpatrick1999, Cardelli1989} used the parameter $R_{\textnormal{V}} = A(V)/(A(B)-A(V))$ for characterizing extinction functions, based on the observation that one parameter would capture most of the variation in extinction across the sky. 

This is a simplifying assumption that holds in certain wavelength regions and breaks down in the UV, where the complexity of the extinction has been shown to be too great for a single parameter \citep{Peek2013}. In this work, for the extinction we consider the wavelength range $0.4$ - $4.5\mu$m, for which one parameter is sufficient to describe the variation in $R_{\textnormal{V}}$, but see \S \ref{sec:extinction_data}. 

Dust grains are heated by absorption of the ambient radiation field and then radiate in the far infrared and microwave. This emission is a major contributor to the foreground of the cosmic microwave background experiments. \cite{Reach1995} and \cite{Finkbeiner1999} made a comprehensive attempt using Far Infrared Absolute Spectrophotometer (FIRAS) and Diffuse Infrared Background Experiment (DIRBE) data to estimate what the contribution from dust is. This was used and improved by the  Wilkinson Microwave Anisotropy Probe (WMAP) team \citep{Bennett2003}, followed by the Planck satellite \citep{Collaboration2014a}.

\cite{Schlafly2016}, hereafter S16, mapped the variation of the dust extinction curve toward different directions on the sky (using tens of thousands of stars), and found a correlation between $R_{\textnormal{V}}$ and the far infrared dust emissivity power law, $\beta$. The emissivity data was obtained from  the Planck satellite \citep{Collaboration2016a}.
We will use the reddening law from S16 to constrain the dust. It is in good agreement with the commonly used \cite{Fitzpatrick1999} reddening curve, including its variation about the mean.  We use S16 because it provides error bars at each of 10 wavelengths providing an obvious way to compute a likelihood, whereas \cite{Fitzpatrick1999} does not.  

Our goal in this analysis is to see if variations in dust grain size distributions and composition can explain the  observed correlation between $R_{\textnormal{V}}$ and $\beta$.
\cite{Weingartner2001a} (hereafter WD01) fit the size distributions using information for the volume of the grains, extinction $A(\lambda)$, and optical parameters of \cite{Laor1993}, under the assumption of spherical grains. In addition to this information, we take into account the $R_{\textnormal{V}}-\beta$ relation found by S16, as well as their reddening law for values ranging between 0.5 and 4.5 $\mu$m (Fig. \ref{fig:range}).  As a result, we can fit a size distribution with the new $R_{\textnormal{V}}-\beta$ constraints and ask what drives the $R_{\textnormal{V}}-\beta$ relation. We start from the parameters describing the size distribution of the grains of dust. The scope is to see if the variation in the 11 parameters of the proposed size distribution function can  explain the  observed correlation between $R_{\textnormal{V}}$ and $\beta$.  We can also explore the effect of a having dust grains exposed to different ISRF intensities. 

In \S \ref{sec:modeling} we explain the modeling for the interstellar dust: its size distribution, composition, extinction, and emission.  In \S \ref{sec:methods} we summarize the optimizer and Markov Chain Monte Carlo method used to constrain the dust size distributions to the reddening vector. Finally, we present our results in \S \ref{sec:results_and_discussion}, and the conclusion in \S \ref{sec:conclusion}.

\section{Modeling} \label{sec:modeling}

\begin{figure}[t]
	\includegraphics[scale=1]{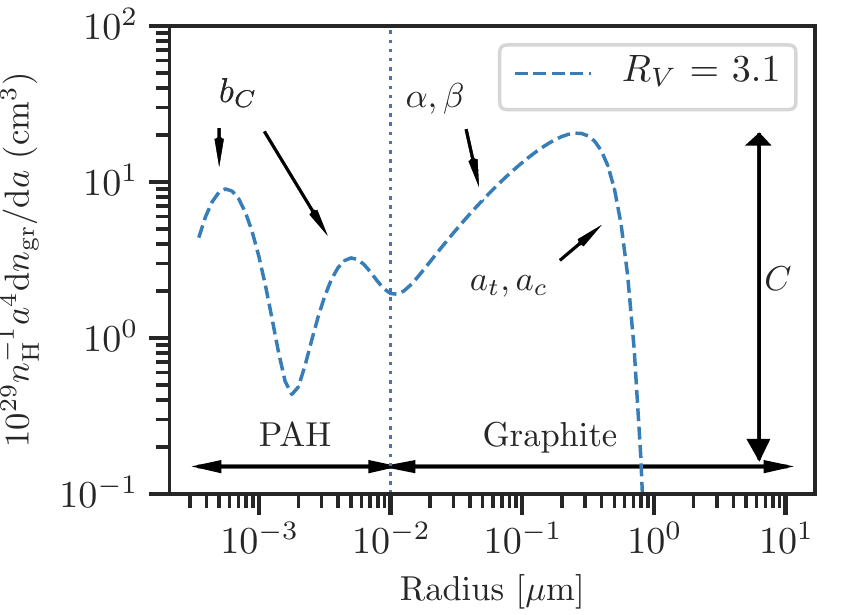}
	\caption{Example of a size distribution for  carbonaceous grains, showing the impact of each of the parameters in the model: $C$ is an overall factor, related to the abundance of carbon atoms per hydrogen nucleus; the exponential ${\alpha}$ in the power-law term $(\frac{a}{a_t})^{\alpha}$ can adjust the slope in  $\frac{\dd \ln n}{ \dd \ln a}$ for $a <  a_t$; $\beta$  can add a positive (for $\beta>0$) or negative (for $\beta<0$) curvature to the slope.  For $a>a_t$, the term $\exp{[-[(a-a_{t})/a_{c}]^3]}$ creates an exponential cutoff whose sharpness can be controlled by $a_c$. In addition, in the function $D(a)$, the sum of two log-normal size distributions for small radius is controlled by its amplitude $b_{\textnormal{C}}$, which represents the total carbon abundance per hydrogen nucleus in the log-normal population. Grains with radii smaller than $10^{-2}\mu$m are modeled as PAHs.\label{fig:size_dist}}
\end{figure}

Our goal is to determine whether variations in the size distributions of the grains of dust can explain the correlation between $R_{\textnormal{V}}$ and $\beta$ found in S16. We use models for the size distributions for different types of grains. Using these models together with models for the absorption and scattering cross sections of the grains,  we are able to calculate the extinction.
Also, together with emission cross section, and an ISRF we can compute an equilibrium temperature for each size and type of grain. Using that, we can predict the collective emission from any size distribution. As a result, we can use these models to study both absorption and emission of dust.

\subsection{Properties of the Dust Grains}

\paragraph{Dust Grain Size Distribution}
We use the models for the dust grain size distributions proposed by WD01 (for work leading up and related to this, see also \cite{Mathis1977},  \cite{Greenberg1978}, \cite{Cardelli1989}, \cite{Desert1990}, \cite{Li2001}, \cite{Li1997} and \cite{Jones2013} for the core-mantle model dust size distribution, and \cite{Wang2015} or the updated version of silicate-graphite model with the addition of a population of large, micron-sized dust grains). An alternative model for the size distribution has been proposed by \cite{Zubko2004}, which can be explored in a future work.
   
In the WD01 model, the dust is modeled using two separate grain populations: silicate composition, and graphite (carbonaceous) composition. For the small carbonaceous grains (radii smaller than $10^{-2}\mu$m), different optical coefficients are used, corresponding to neutral and ionized polycyclic aromatic hydrocarbons (PAHs). PAHs are structures made of hexagonal rings of carbon atoms with hydrogen atoms attached to the boundary. It is assumed that neutral and ionized PAHs each give half of the contribution of the PAHs. Other types of grains (such as oxides of silicon, magnesium, and iron, carbides, etc.) are not included.

The size distributions are modeled by Eq. \ref{eq:size1} and \ref{eq:size2}.

\begin{widetext}
	\begin{equation}\label{eq:size1}
	\frac{1}{n_{\textnormal{H}}}\dv{n_{\textnormal{gr}}(a)}{a}= D(a) + \frac{C}{a}\Big(\frac{a}{a_{t}}\Big)^{\alpha_g} \times
	\left\{
	\begin{array}{ll}
	1+ \beta a/a_t, & \beta \geq 0 \\
	(1-\beta a/a_t)^{-1}, & \beta < 0
	\end{array}
	\right\}    
	\times
	\left\{
	\begin{array}{ll}
	1, & \text{3.5 \AA }< a< a_{t}  \\
	\exp{[-[(a-a_{t})/a_{c}]^3]}, & a>a_{t}
	\end{array}
	\right.
	\end{equation}
	\begin{equation}\label{eq:size2}
	D(a) = 
	\left \{
	\begin{array}{ll}
	0, & \text{for silicate dust}\\
	2.04 \cdot 10^{-2} \frac{ bC}{a}e^{-3.125 ( \ln{(a/3.5 \text{\AA})})^2} + 9.55\cdot 10^{-6} \frac{ bC}{a} e^{-3.125 ( \ln{(a/30 \text{\AA})})^2}, &\text{for carbonaceous/PAH dust}
	\end{array} 
	\right.
	\end{equation}
\end{widetext}


These equations are described by 11 parameters: five corresponding to the silicate population and six to the PAH and graphite (Fig. \ref{fig:size_dist}).

\paragraph{Optical Parameters of the Dust Grains}
To calculate the emission and extinction of dust we need to know the optical properties of the  grains of dust, such as the absorption and scattering coefficients. For silicates and graphite, we use the values derived by \cite{Laor1993}  and \cite{Draine1984}. For PAH-carbonaceous grains we use the properties obtained in \cite{Li2001}\footnote{The files that were used in this analysis can be found on Professor Bruce Draine's website \url{https://www.astro.princeton.edu/~draine/dust/dust.diel.html}. The specific files are files Gra\_81.gz, PAHion\_30.gz, PAHneu\_30.gz and Sil\_81.gz.}. 

The models contain 81 log-spaced radii between $10^{-3}\mu$m and $10 \mu$m for silicates,  30 from $3.55\times10^{-4}\mu$m to $10^{-2}\mu$m  for PAHs, and 61 between $10^{-2}\mu$m and $10 \mu$m for graphite \footnote{The file for graphite, Gra\_81.gz, actually has 81 log-spaced radii between $10^{-3}\mu$m and $10 \mu$m, but we use only the 61 between $10^{-2}\mu$m and $10 \mu$m to complement the range of radius for the PAHs.}. 

\cite{Laor1993} model dust grains as solid spheres of radius $a$ with absorption cross section at wavelength $\lambda$ of $C_{\textnormal{abs}}(\lambda,a) $. They label the scattering cross section at wavelength $\lambda$  with
$C_{\textnormal{sca}}(\lambda, a) $ and the extinction cross section with $C_{\textnormal{ext}}(\lambda, a) \equiv C_{\textnormal{abs}}(\lambda, a) + C_{\textnormal{sca}} (\lambda, a) $. The scattering and absorption efficiencies $Q_{\textnormal{sca}}$ and $Q_{\textnormal{abs}}$ are defined as:

\begin{equation}
Q_{\textnormal{sca}}(\lambda, a) \equiv \frac{C_{\textnormal{sca}}(\lambda, a)}{\pi a^2}; Q_{\textnormal{abs}}(\lambda, a) \equiv \frac{C_{\textnormal{abs}}(\lambda, a)}{\pi a^2}.
\end{equation}

The wavelength range for the optical parameters for all types of grains is $10^{-3}\mu$m to $1$mm. The graphite and silicate files have 241 log-spaced wavelength samples. For the PAH files, their wavelength array is five times more dense than the wavelength array from the graphite or silicate files, so we take only every fifth value, corresponding to exactly the same values as the sampling of the graphite and silicate files.

\begin{figure}[t]
	\includegraphics[scale=1]{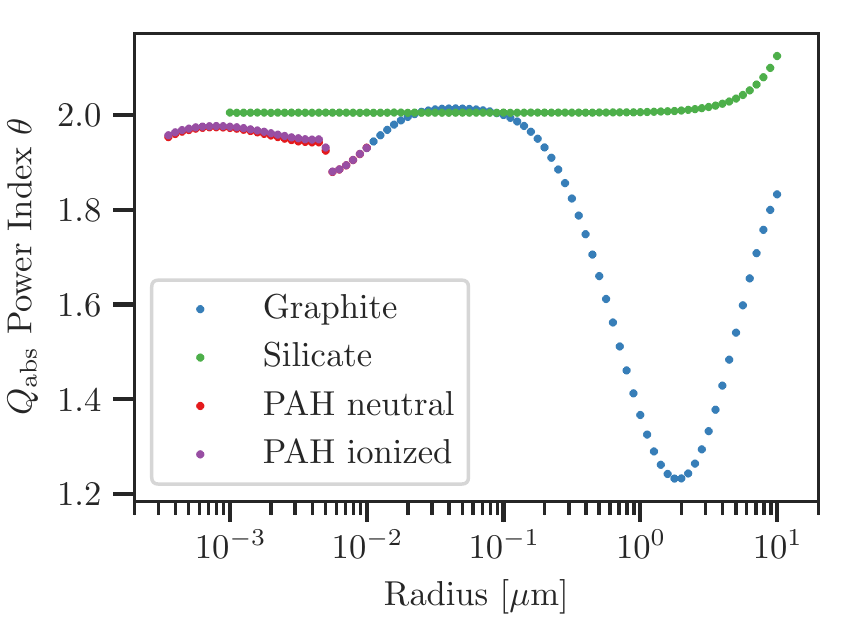}
	\caption{The absorption optical coefficients for carbonaceous and silicate grains can be approximated with a power law for the wavelength range between $335$ and $1000\mu$m. The power-law index has a different dependence on radius for each type of grain. In a collection of grains, carbonaceous grains can contribute higher $\theta$s than the silicate grains, leading to lower $\beta$ index for the entire collection. \label{fig:power_index}}
\end{figure}

For the wavelength range between $335\mu$m and $1000\mu$m, we model the absorption using a power law, $Q_{\textnormal{sca,abs}}(\lambda,a) = \tau(a)\cdot({\lambda/\lambda_0})^{-\theta(a)}$ (Appendix \ref{sec:extend}).
We are interested in looking at the absorption coefficient behavior for different compositions (Fig. \ref{fig:power_index}). What we notice is that carbonaceous and silicate grains show quite a different power law index as a function of the radius. Thus, one can expect to control the resulting emissivity power-law index $\beta$ for a collection of dust grains by changing composition or size. We use the power-law fit of the absorption optical properties to extend them to $10^4\mu$m. This range is more in line with future cosmic microwave background (CMB) experiments such as The Primordial Inflation Experiment \citep[PIXIE]{Kogut2011}.

\subsection{Extinction}\label{sec:extinction}

\subsubsection{Extinction Modeling}\label{sec:extinction_modeling}
For a given collection of dust grains along the line of sight, we want to calculate the extinction $A$, defined as:
\begin{equation}
A(\lambda) = m_{\textnormal{attenuated}} - m_0 = 2.5\log_{10}\frac{F^0_{\lambda}}{F_{\lambda}^a}
\end{equation}
where $F_{\lambda}^a ,\: m_{\textnormal{attenuated}} $ are the dust attenuated observed flux and magnitude of the object, and  $F^0_{\lambda},\:  m_0$ are the flux and magnitude that would have been observed if there would have been no attenuation from dust. Thus, extinction can be related to the optical depth $\tau(\lambda)$ by $A(\lambda) = (2.5 \log_{10}e) \tau(\lambda)$. The optical depth is created from the contributions of each grain along the line of sight. Let $i$ be the index of the grain type, referring to graphite, PAHs, or silicates. For grains of radius $a$ of type $i$, their impact on the optical depth can be expressed as the product of an effective extinction cross section $C_{\textnormal{ext},i}(\lambda, a) $ and the column density $N_{i}(a)$. Then, the optical depth given by a distribution of grains of different radii $a$ is given by
\begin{equation}
\tau (\lambda)= \sum_i \int \dv{N_i}{a} C_{\textnormal{ext},i}(\lambda, a)\dd{a}
\end{equation}

The fraction of dust grains per radius becomes:
\begin{equation}
\dv{N_i(a)}{a} = \dv{}{a}\int n_i(a,s) \dd s =\dv{}{a}\int \left(\frac{n_i}{n_{\textnormal{H}}}\right) (a,s)n_{\textnormal{H}}(s) \dd s ,
\end{equation}
where $s$ is the path length along the direction of integration, $n_i$[grains cm$^{-3}$] is the number of dust particles of type $i$ per volume, $n_{\textnormal{H}}$[atoms cm$^{-3}$] is the number of hydrogen atoms per volume, and $N_{\textnormal{H}}$[atoms cm$^{-2}$] $= \int  n_{\textnormal{H}}(s) \dd s$ is the hydrogen column density. In this analysis we assume the dust to gas ratio is constant along the line of sight $s$. As a result,
\begin{equation}
\dv{N_i(a)}{a} = \left(\int n_{\textnormal{H}}(s) \dd s \right ) \frac{1}{n_{\textnormal{H}}}\dv{n_i(a)}{a}= \frac{N_{\textnormal{H}}}{n_{\textnormal{H}}} \dv{n_i(a)}{ a}.
\end{equation}
Using the fact that $\dd{a} = a \dd{\log{a}}$, the optical depth can then be calculated as:
\begin{equation}
\frac{\tau (\lambda)}{N_{\textnormal{H}}} = \pi \sum_i \int \frac{1}{n_{\textnormal{H}}} \dv{n_i(a)}{a} Q_{\textnormal{ext},i}(\lambda,a)a^3 \dd{\log{a}}.
\end{equation}

The extinction $A_{\lambda}$ over the column density is:

\begin{equation}
\label{eq:11_parameters_extinction_equation}
\frac{A (\lambda)}{N_{\textnormal{H}}} = (2.5 \log_{10}e) \pi \sum_i \int \frac{1}{n_{\textnormal{H}}} \dv{n_i(a)}{a} Q_{\textnormal{ext},i}(\lambda,a)a^3 \dd{\log{a}}
\end{equation}

\begin{table}[!]
\centering
\begin{tabular}{llllll}
\hline
Filter  & g  & r & i & z & y \\
\hline
$\lambda[\mu]$m & 0.503 & 0.6281 & 0.7572 & 0.8691 & 0.9636  \\
\hline
$\nu$[THz] & 595.8 & 477.3 & 395.9 & 344.9 & 311.1 \\
\hline
Filter  & J  & H & K & W1 & W2 \\
\hline
$\lambda[\mu]$m &  1.2377  & 1.6382 & 2.1510 & 3.2950 & 4.4809  \\
\hline
$\nu$[THz] & 242.2 & 183.0 & 139.4 & 90.98 & 66.90 \\
\hline

\end{tabular}
\caption{Wavelength and frequency values for the ten points where we compare the modeled extinction with the extinction data coming from S16 reddening vector.\label{table:extinction_table}}
\end{table}

\subsubsection{Extinction Data}\label{sec:extinction_data}

S16 derived the dust extinction curve towards 37,000 stars in different directions across the sky. Using photometry from Pan-STARRS1 \citep{Hodapp2004,Chambers2016}, Two Micron All-Sky Survey \citep[2MASS,][]{Skrutskie2006} and the Wide-field Infrared Survey Explorer \citep[WISE,][]{Wright2010,Cutri2013}, and spectra from the APOGEE survey \citep{Majewski2017,Eisenstein2011}, they performed a principal component analysis and found that the extinction function is well approximated by two principal components, called the vector $\bm{R}_0$ (constant across the directions in the sky) and a perturbation vector $\dv{\bm{R}}{x}$. Both $\bm{R}_0$ and $\dv{\bm{R}}{x}$ have norm 1. The extinction function can be expressed as:

\begin{equation}
\bm{A}_{\textnormal{Schlafly}} = \bm{R_0} + x \dv{\bm{R}}{x},
\label{eq:schlafly_extinction}
\end{equation}
where $x$ is a parameter that varies across the sky, with values between -0.4 and 0.4. Extinction laws are usually characterized by the parameter $R_{\textnormal{V}} =\frac{A(\textnormal{V})}{A(\textnormal{B})-A(\textnormal{V})}$. However, since S16 did not have access to the distances to the star, the absolute gray component of the extinction is not known. Instead, they approximate the $R_{\textnormal{V}}$ parameter with $ R_{\textnormal{V}}' = 1.2 \frac{A(g)-A(W2)}{A(g)-A(r)}-1.18$. The parameter $x$  is related to $ R_{\textnormal{V}}'$ using equation \ref{eq:x_to_Rv}:

\begin{equation}\label{eq:x_to_Rv}
R_{\textnormal{V}}' = 3.3 + 9.1 x
\end{equation}

The intent was that $x=0$ ($R_{\textnormal{V}}'$ of 3.3) corresponds to a mean reddening vector. However, this results in an $R_{\textnormal{V}}'$ of 3.3 at \cite{Fitzpatrick1999} $R_{\textnormal{V}}$ of 3.1. Subsequently, in this analysis, we use the notation $R_{\textnormal{V}}$ to refer to the $R_{\textnormal{V}}'$ from S16.

The reddening vector is specified at the wavelengths/frequencies showed in Table \ref{table:extinction_table}. These wavelengths/frequencies have been obtained by S16 by weighting over the M-giant star spectrum and over the bandpass of the detectors. $\nu_{\textnormal{mean},b} = \frac{\int \nu S_{\nu} F_{\nu, b} \dd{\nu}}{\int S_{\nu} F_{\nu, b} \dd{\nu}}$, where $S_{\nu}$ is the M-giant spectrum, 
$b$ represents the index of the band (g, r, i, \dots), and $F_{\nu, b}$ represents the filter weight.

\cite{Li2014} found that the aliphatic 3.4 $\mu$m C-H stretch absorption band is seen in diffuse clouds, and absent in dense regions. Therefore, for lines of sight with larger $R_{\textnormal{V}}$, the 3.4 $\mu$m extinction band is weaker or even absent. This raises the question of whether  $R_{\textnormal{V}}$-based \cite{Cardelli1989} parameterization is valid only at $\lambda < 3\mu$m. However, within the range of $E(\textnormal{B}-\textnormal{V})$ measured in S16, they did not see evidence for significant variation in 3.4 $\mu$m W1 and 4.6 $\mu$m W2 Wide-field Infrared Survey Explorer \cite{Wright2010} bands. Since in this work we employ the extinction laws of S16, the $R_{\textnormal{V}}$ parameter is used.

\subsection{Grain equilibrium temperature as a function of radius}\label{sec:equilibrium_temp}
\begin{figure}[t]
	\includegraphics[scale=1]{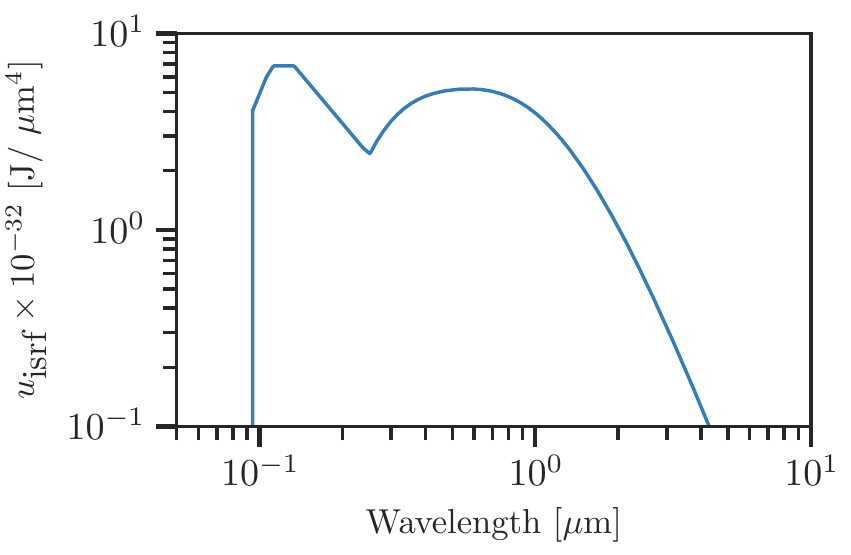}
	\caption{Mean interstellar radiation field from \cite{Mezger1982} and \cite{Mathis1983}, for $\chi_{\textnormal{ISRF}}$ = 1.\label{fig:isrf}}
\end{figure}

To calculate the thermal radiation emitted by a collection of dust grains, we need to know the temperatures of the grains.  The grains are exposed to the ambient radiation field. In this calculation, we take into account only radiative heating and ignore the collisional heating that would be provided to the grains in the situation when they are surrounded by gas. In the case of dense clouds, however, this can become a relevant contribution.

\paragraph{Interstellar Radiation Field} 
We follow \S 4 of \cite{Weingartner2001b} and use the interstellar radiation field (ISRF) model of \cite{Mezger1982} and \cite{Mathis1983}. The radiation field as a function of frequency (Fig. \ref{fig:isrf}) is:

\begin{widetext}

\begin{equation}\label{eq:isrf}
\nu u_{\nu}^{\textnormal{\textnormal{ISRF}}} =  \left \{
	\begin{aligned}
    &0 &&,\ h\nu > 13.6 \textnormal{eV} \\
	&3.328 \times 10^{-9}\textnormal{erg  \ cm}^{-3}(h \nu / \textnormal{eV})^{-4.4172}  &&,\  11.2 < h\nu <13.6 \textnormal{eV} \\
    &8.463 \times10^{-13}\textnormal{erg  \ cm}^{-3}(h \nu / \textnormal{eV})^{-1} &&,\ 9.26 < h\nu <11.2 \textnormal{eV} \\
    &2.055 \times10^{-14}\textnormal{erg  \ cm}^{-3}(h \nu / \textnormal{eV})^{0.6678} &&,\ 5.04 < h\nu <9.26 \textnormal{eV} \\
    &(4\pi\nu/c)\sum^3_{i=1}w_iB_{\nu}(T_i) &&,\ h\nu<5.04\,\textnormal{eV}
    \end{aligned} \right.
\end{equation}

\end{widetext}
where $w_1=1\times10^{-14}, w_2 =1.65\times 10^{-13} ,w_3 = 4 \times 10^{-13}$, and $T_1 = 7500$K, $T_2 = 4000$K, $T_3 = 3000$K.
In our analysis, we would want to modify the radiation field to account for inhomogeneities in the interstellar medium, where we can have areas that are hotter than others. For that, as an approximation, we will multiply the radiation field by a factor $\chi_{\textnormal{ISRF}}$ that varies from 0.5 to 2. 

\paragraph{The thermal equilibrium equation}

\begin{figure}[t]
\includegraphics[scale=1]{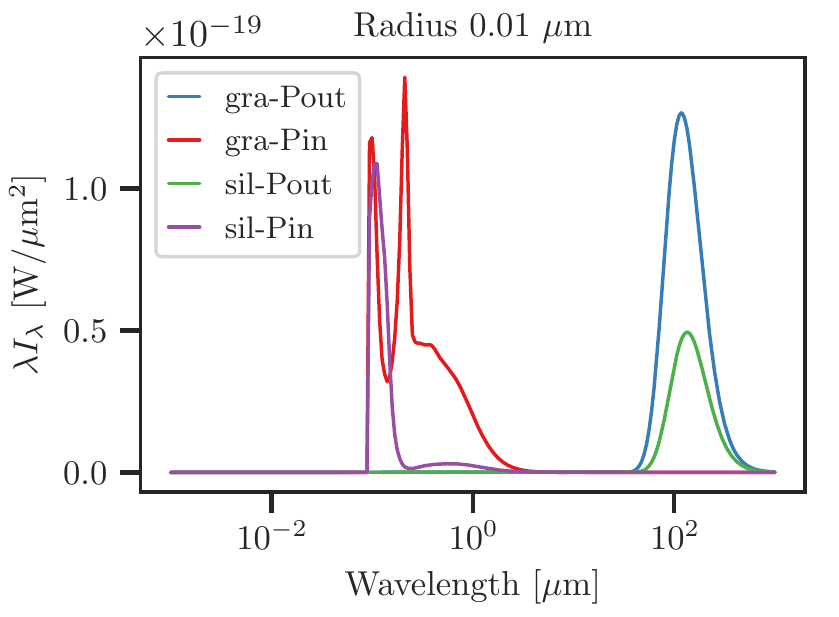}
\caption{Power input and output for a single grain of radius 0.01$\mu$m. $\lambda I_{\lambda}$ is the power per $\log\lambda$. ie. the area under the curve is the power. This shows that the power absorbed equals the power emitted, for the case of equilibrium temperature.  \label{fig:pin_pout}}
\end{figure}
For each grain radius $a$, we  assume thermal equilibrium between the absorbed radiation and emitted radiation ($P_{\textnormal{in}} = P_{\textnormal{out}}$, Fig. \ref{fig:pin_pout})  We assume the grain is spherical and emits like a black body of unknown temperature $T$, which we aim to determine. The absorbed radiation is assumed to come from the interstellar radiation field surrounding the grain sphere uniformly.

The thermal equilibrium equation for one dust grain of size $a$ is thus:

\begin{equation}
\begin{split}
\int_0^{\infty} Q_{\textnormal{abs}}(\lambda,a)\pi a^2\chi_{\textnormal{ISRF}} u_{\textnormal{ISRF}}(\lambda)\dd\lambda = \\\int_0^{\infty} Q_{\textnormal{abs}}(\lambda,a)\pi a^2 \frac{4\pi}{c} B_{\lambda}(T)\dd\lambda
\end{split}
\label{eq:thermal_equilibrium}
\end{equation}

\begin{figure}[t!]
\includegraphics[scale=1]{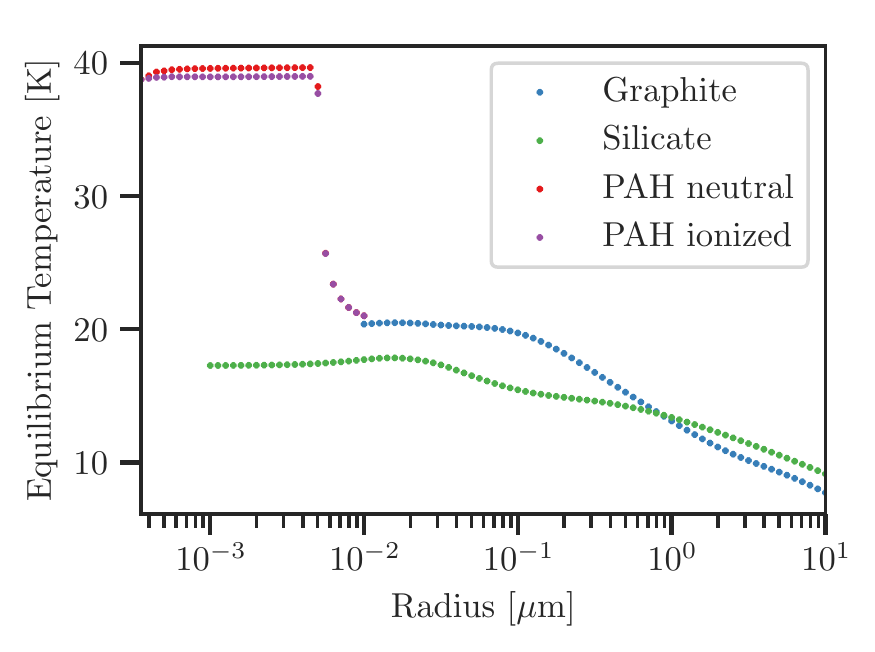}
\caption{The calculated equilibrium temperatures for the 4 types of grains for different values of radii, for $\chi_{\textnormal{ISRF}}$ = 1.\label{fig:equilibrium_temperature}}
\end{figure}

The integral in  Eq. \ref{eq:thermal_equilibrium} is taken over the wavelength range from $10^{-3}\mu$m to $10$mm, using the extension shown in Appendix \ref{sec:extend}. We obtain the equilibrium temperatures for 4 types of grains for different radii (Fig. \ref{fig:equilibrium_temperature}).

This method of calculation for the equilibrium temperature assumes that at each radius for each type of grain there is a single temperature. This approximation breaks down as the radius of the grain becomes small enough. A grain stays at an equilibrium temperature if no one photon it absorbs or emits carries enough energy to perturb the temperature much. Big grains have a thermal energy much larger than one photon. But small grains do not: a single photon with several eV carries more energy than the entire thermal energy of the grain, and the emission and absorption of a single quanta can create temperature spikes. In our calculation, we are considering grains as small as $3.55 $\AA; especially for grains smaller than $10^{-2}\mu$m (like the PAHs), in a future study, there can be a benefit from replacing the approximation with a different method  where one considers a temperature distribution for each radius size, as done in the work of \cite{Draine2001}. However, in this work we are mainly interested in long wavelength emission where $\langle B_{\nu}(T)\rangle = B_{\nu}(\langle T\rangle )$.

\begin{figure}[t]
	\includegraphics[scale=1]{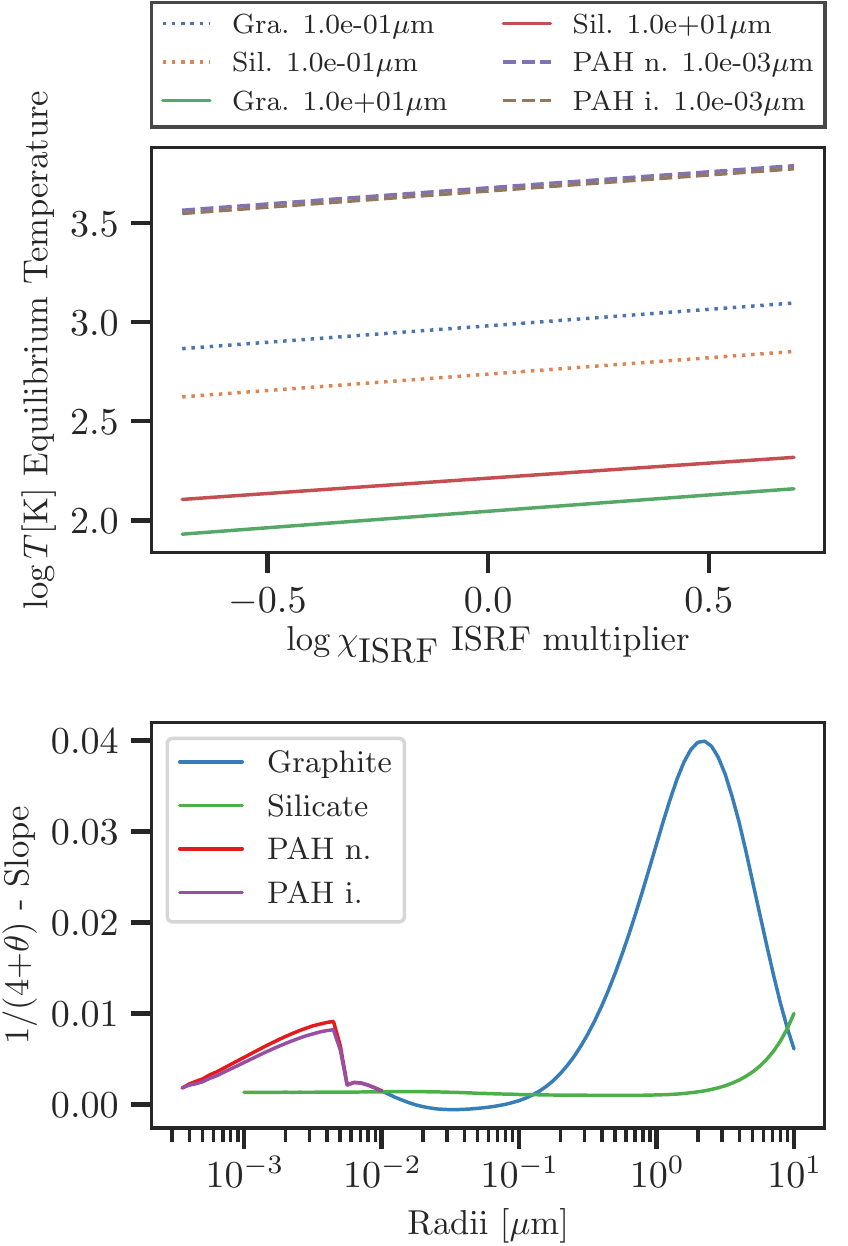}
	\caption{Top panel: log of the equilibrium temperature ($T$) as a function of the log of the multiplier of the interstellar radiation field ($\chi_{\textnormal{ISRF}}$), for carbonaceous, silicate, and PAH (neutral and ionized) grains at select radii. They follow a linear dependence. The temperature curves are calculated at fixed values for the radius of the grains, as specified in the legend. Bottom panel: the difference between $1/(4+\theta)$ ($\theta$ refers to the power-law indexes of the absorption optical coefficients as seen in Fig. \ref{fig:power_index}) and the slopes of the linear fits versus grain radii, for each type of grain. The difference is significant to warrant not using the $1/(4+\theta)$ approximation, and it serves as a good check for our calculations. \label{fig:gamma_temperature}}
\end{figure}

In addition to the grain size and type, the variation of the ISRF throughout the ISM leads to variations in temperature. To reproduce this effect, we allow the ISRF multiplier parameter $\chi_{\textnormal{ISRF}}$ to vary between 0.5 and 2, and calculate the equilibrium temperature for the range (Fig. \ref{fig:gamma_temperature}).

\subsection{Modeling Emission}\label{sec:modeling_emission}
\subsubsection{Calculating the emission intensity from a collection of grains}

The emissivity (power radiated per unit volume
per unit frequency per unit solid angle) coming from a collection of grains is defined as:
\begin{equation}
j_{\nu}=\sum_i \int \dd a \dv{n_i}{a}C_{\textnormal{abs},i}(\nu,a)B_{\nu}(T_{\textnormal{eq}}(a))
\end{equation}
where $i$ = the index for carbonaceous, silicate, and PAH grains.
The spectral intensity $I_{\nu}$ is defined as the emissivity integrated along the line of sight $s$:
\begin{equation}
I_{\nu} = \int j_{\nu} \dd s  = \int \left(\frac{j_{\nu}}{n_{\textnormal{H}}}\right)(s) n_{\textnormal{H}}(s) \dd s 
\end{equation}
Since $\frac{n_i}{n_{\textnormal{H}}}$ is assumed to be constant along the line of sight, $\frac{j_{\nu}}{n_{\textnormal{H}}}$ becomes constant along the line of sight as well. Using $ N_{\textnormal{H}} = \int  n_{\textnormal{H}}(s) \dd s$ , we obtain:

\begin{equation}
\begin{split}
I_{\nu} & =  \frac{j_{\nu}}{n_{\textnormal{H}}} \int  n_{\textnormal{H}}(s) \dd s = \frac{j_{\nu}}{n_{\textnormal{H}}} N_{\textnormal{H}}   = \\
& = N_{\textnormal{H}} \sum_i \int \dd a \frac{1}{n_{\textnormal{H}}}\dv{n_i}{ a} C_{\textnormal{abs},i}(\nu,a)B_{\nu}(T_{\textnormal{eq}}(a))
\end{split}
\end{equation}

\subsubsection{The Modified Black Body Fit}\label{sec:MBB_fit}

We aim to compare our analysis with the 2013 Planck release \citep{Collaboration2014b}\footnote{The spectral index data can be found in the file HFI\_CompMap\_ThermalDustModel\_2048\_R1.20.fits at \url{https://irsa.ipac.caltech.edu/data/Planck/release_1/all-sky-maps/previews/HFI_CompMap_ThermalDustModel_2048_R1.20/index.html}. We select the directions in the sky to reproduce the same analysis done by S16, whose data is available at \url{https://dataverse.harvard.edu/dataset.xhtml?persistentId=doi:10.7910/DVN/WMA5KJ}. The HEALPIX binning used was NSIDE=64.}, as was used by S16 . 

Spectral energy density (SED) of emission from dust has been modeled in practice by the \cite{Collaboration2014b} using a modified black body (MBB) function:
\begin{equation}\label{eq:MBB}
I_{\nu} = \tau_{353} \big ( \frac{\nu}{353\text{ GHz}} \big )^{\beta} B_{\nu}(T), 
\end{equation}
where $ B_{\nu}(T)$ is the Planck function for dust of temperature $T$. $353$GHz is chosen as a reference frequency. The assumption is used in the optically thin limit. The power law parameterized by $\beta$ models the dependence of the emission cross-section with frequency. The fit for the three parameters in Equation \ref{eq:MBB} is performed using data from four photometric bands: 353GHz, 545GHz, 857GHz from Planck, and 3000GHz ($100 \mu$m) from IRAS \citep{Schlegel1998,Beichman1988}. Because these are the bandpasses the \cite{Collaboration2014b} used in their analysis, to compare to their results, we evaluate the intensity at the same four bandpasses. We use the weighting given in Appendix B/Table 1 of \cite{Collaboration2014b}, and the corresponding response functions \footnote{The Planck filter files can be found on the website  \url{http://pla.esac.esa.int/}, in the section called "Software, Beams, and Instrument Model". At the time of this paper, Planck has 3 releases HFI\_RIMO\_R1.10.fits (2013), HFI\_RIMO\_R2.00.fits (2015), HFI\_RIMO\_R3.00.fits (2016). We use HFI\_RIMO\_R1.10 fits because it was the one used for the data release from the \cite{Collaboration2014b}. For IRAS, the filter files can be found on the website https://irsa.ipac.caltech.edu/IRASdocs/exp.sup/ch2/tabC5.html}. 


\setlength{\tabcolsep}{4pt} 

\begin{table*}[t]
\centering
\begin{tabular}{llllllllllll}
\toprule\toprule
Parameter  &$10^5$ $b_{\textnormal{C}}$ & $\alpha_g$ & $\beta_g$ & $a_{t,g}[\mu \textnormal{m}]$ & $a_{c,g}[\mu \textnormal{m}]$ & $10^{27}V_g [\textnormal{cm}^3$ $\textnormal{H}^{-1}]$& $\alpha_s$ & $\beta_s$ & $a_{t,s}[\mu \textnormal{m}]$ & $a_{c,s}[\mu \textnormal{m}]$ & $10^{27}V_s [\textnormal{cm}^3$ $\textnormal{H}^{-1}]$    \\ \hline
Lower Boundary & 3.0 & -3.0  & -30. &0.000355 & 0.000355 & max $b_{\textnormal{C}}$ & -3.0 & -30. & 0.001 & 0.001 & 0.1                  \\ 
\hline
Upper Boundary & 5.0-6.5 & -0.5  & 30.  &10.00000 & 10.00000 & 6$\times$2.07=12.42     & -0.5 & 30.  & 10.00 & 10.00 & 6$\times$2.98 = 17.88\\
\bottomrule \bottomrule
\end{tabular}
\caption{The boundaries for the parameter space explored by the MCMC. 
	Since $a_t$ controls the position of the exponential drop, we allow it to values over the range of the grains. $a_c$ controls the smoothness of the exponential factor, so it should be able to get values of similar magnitude to $a_t$. As such, we give it the same range. Gaussian priors are given for the carbonaceous ($V_g$=PAH+graphite) and silicate ($V_s$) volumes that are centered within the range. The $V_g$ parameter's low bound is set to be large enough to account for the maximum possible contribution coming from Eq. \ref{eq:size2} with the highest allowed $b_{\textnormal{C}}$ value. The maximum values for $V_g$ and $V_s$ are set to be 6 times the reference values in WD01. The limits on the $b_{\textnormal{C}}$ parameter are explained in \S \ref{sec:MCMC_analysis_results}. 
	\label{table:fiducial_table}}
\end{table*}
\section{Methods} \label{sec:methods}

The goal of this work is to explore the space of WD01 grain size distributions to find those that are consistent with our prior knowledge about dust, including:
\begin{enumerate}
\item the \emph{shape} of the reddening curve and its variation with $R_{\textnormal{V}}$,
\item the \emph{amount} of reddening per H atom, and
\item the abundance of metals (C, Si, etc.) per H atom required to make dust.
\end{enumerate}
For each sample from the WD01 parameter space, we compute the emission spectrum expected for dust in a reference radiation field, and fit the $\tau$, $\beta$, and $T$ parameters of a modified black body (MBB) as described in \S \ref{sec:modeling_emission}.  Combining the emission and extinction for each sample, we can study the relation between the $R_{\textnormal{V}}$ and $\beta$ parameters.

To perform our analysis, we create a reference extinction function, constrain the gray component of the extinction, normalize to physical values of extinction per column density of hydrogen ($A(I)/N_{\textnormal{H}})$, impose physical boundaries on the parameters, and keep the total mass per H atom of the grain distributions within an expected range.

An MCMC-like algorithm is used to explore the available space for the size distribution parameters. However, MCMCs are not an optimization technique, and they can take a long time to converge on the points of high likelihood. 
To help the solver converge, an optimizer is run to obtain an initial guess for the MCMC. A possible downside is that by doing this, a region of the parameter space might remain unexplored.

Using the posterior points from the MCMC analysis we create the emissivity corresponding to the Planck bandpass filters and perform a modified black body fit as described in section \S \ref{sec:modeling_emission}.

Using the data for extinction and emissivity, we explore the relation between the $R_{\textnormal{V}}$ and $\beta$ parameters.

\paragraph{Creating the reference extinction functions}\label{sec:ref_ext}

S16 focused on the \emph{shape} of the reddening curve by using the relative extinction in 10 bands. Their work does not inform us about the gray component (which could not be measured without knowing the distance to each star) or the overall amplitude per H atom (which they did not measure).  
Therefore our first step is to establish a "target" reddening curve by setting an additive and multiplicative term to fix these degrees of freedom. The condition that $A(H)/A(K)$ = 1.55 \citep{Indebetouw2005} constrains the additive term as follows: we define $\bm{A'}_{\textnormal{Schlafly}} = \bm{A}_{\textnormal{Schlafly}} + C_{\textnormal{Schlafly}} $ with $\bm{A}_{\textnormal{Schlafly}} = \bm{R_0} + x \dv{\bm{R}}{x}$

\begin{equation}
\begin{split}
&\frac{A_{\textnormal{Schlafly}}(H)+C_{\textnormal{Schlafly}}}{A_{\textnormal{Schlafly}}(K) + C_{\textnormal{Schlafly}}} = 1.55  = r\\
&A_{\textnormal{Schlafly}}(H) + C_{\textnormal{Schlafly}} = A_{\textnormal{Schlafly}}(K)r+C_{\textnormal{Schlafly}}r\\
&C_{\textnormal{Schlafly}} = \frac{A_{\textnormal{Schlafly}}(H)-A_{\textnormal{Schlafly}}(K)r}{r-1}  \\
&\bm{A}_{\textnormal{Schlafly}}' = \bm{A}_{\textnormal{Schlafly}} + C_{\textnormal{Schlafly}} 
\end{split}
\end{equation}
Fixing the  grey component creates degeneracy between $ A(H)$ and $A(K)$. To maintain the correct number of degrees of freedom, we remove the $H $ band from the $\bm{A}$ vectors and therefore from the covariance matrix and the $\Delta \chi^2$ calculation. $ A(H)$ is determined by the other parameters and is no longer independent, so it can be ignored in the calculation and recovered at the end.

Having fixed the additive term, we now impose an extinction per N(H) assumption to fix the multiplicative term. \cite{Cardelli1989} suggested the convention that $A(I)/N_{\textnormal{H}} = 2.6\times10^{-22}\textnormal{cm}^2$. To be consistent with the extinction functions presented in WD01, we define $A(I)/N_{\textnormal{H}} = 3.38\times10^{-22}\textnormal{cm}^2$. We denote this quantity by $C_{\frac{A_I}{N_{\textnormal{H}}}}$, and the normalized extinction by $\frac{\bm{A''}_{\textnormal{Schlafly}(\lambda)}}{N_{\textnormal{H}}} = \frac{ \bm{A}_{\textnormal{Schlafly}}'(\lambda)}{ \bm{A}_{\textnormal{Schlafly}}'(I)} \times C_{\frac{A_I}{N_{\textnormal{H}}}}$. 
 $C_{\frac{A_I}{N_{\textnormal{H}}}}$ is a convention, not a measurement with an error. In reality its value most likely varies across the sky. If a future experiment makes a different measurement of $A_{\textnormal{I}}/N_H$, it should be taken into account.
 
 Thus, the reference extinction function can be constructed using:

\begin{equation}
\frac{\bm{A}_{\textnormal{reference}(\lambda)}}{N_{\textnormal{H}}} = \frac{\bm{A''}_{\textnormal{Schlafly}(\lambda)}}{N_{\textnormal{H}}} = \frac{ \bm{A}_{\textnormal{Schlafly}}'(\lambda)}{ \bm{A}_{\textnormal{Schlafly}}'(I)} \times C_{\frac{A_I}{N_{\textnormal{H}}}}
\label{eq:reference_extinction}
\end{equation}

\paragraph{Volume of the dust grains}\label{sec:vol_priors}
As we let the MCMC and the optimizer explore the parameter space, we want to make sure the size distribution does not require more atoms (per H) of C and Si than are available in the universe. As such, we want to have the total mass per H atom of the dust grains as an upper bound in the parameter limits. WD01 phrases this constraint in terms of the volume per H atom, so we use this notation in this analysis.

 Thus, we introduce Gaussian volume priors. For consistency, we center the priors at the values adopted by WD01 for the volume found of each type of grain in the universe, with a standard deviation of 10$\%$. For carbonaceous grains, the total expected volume is $V_{\textnormal{tot},g} \approx 2.07 \times10^{-27}\textnormal{cm}^3$ $\textnormal{H}^{-1}$, and for silicates $V_{\textnormal{tot},s} \approx 2.98 \times10^{-27}\textnormal{cm}^3$ $\textnormal{H}^{-1}$. The mean of the Gaussian in the prior is fixed for now but later in \S \ref{sec:optimizer_results} we will vary it.

$b_{\textnormal{C}}$ represents the overall amplitude of the bumps; to calculate the PAH volumes, one needs to also add part coming from the non-$D(a)$ part of the size distribution, integrated over the range of the PAH radii.

Since $C$ is an overall factor, it can be calculated from a proposed combination of volume $V$ and parameters $\alpha, \beta, a_{t}, $ and $a_{c}$. As a result, we replace the $C$ parameter with a volume parameter $V$, and calculate the corresponding $C$ when needed for the size distribution calculations.

Thus, the 11 parameters explored by the MCMC are $b_\textnormal{C},\ \alpha_g,\  \beta_g,\  a_{t,g},\  a_{c,g},\  V_g,\  \alpha_s,\  \beta_s,\  a_{t,s},\  a_{c,s},\ $and $  V_s$. Table \ref{table:fiducial_table} lists the boundaries for each parameter.

\begin{figure*}[t!]
	\centering
	\begin{tabular}{cccc}
		\hspace{-5.00mm}
		\includegraphics[width=0.35\textwidth]{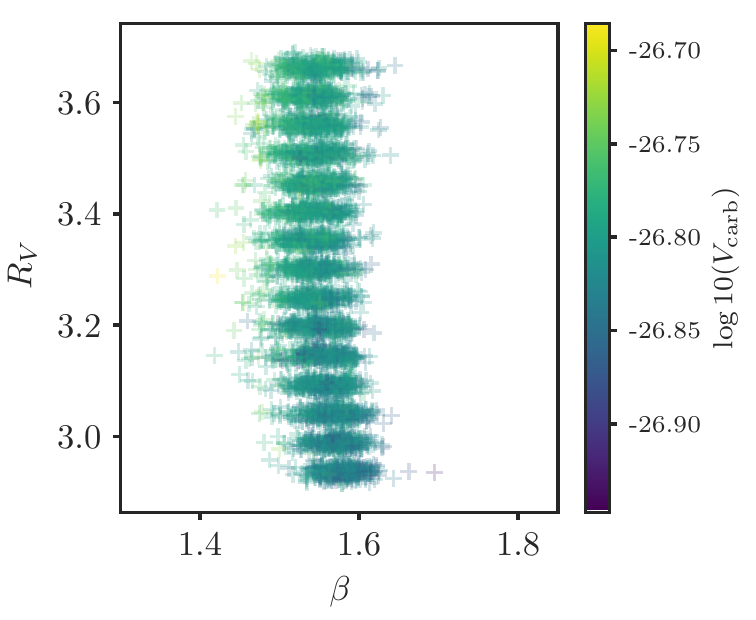} &
		\hspace{-5.00mm}	
		\includegraphics[width=0.35\textwidth]{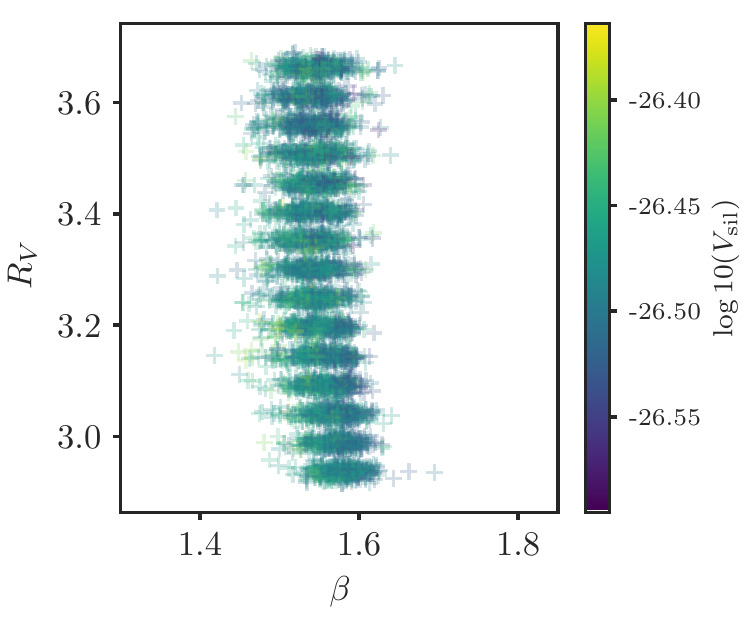} &
		\hspace{-5.00mm}
		\includegraphics[width=0.35\textwidth]{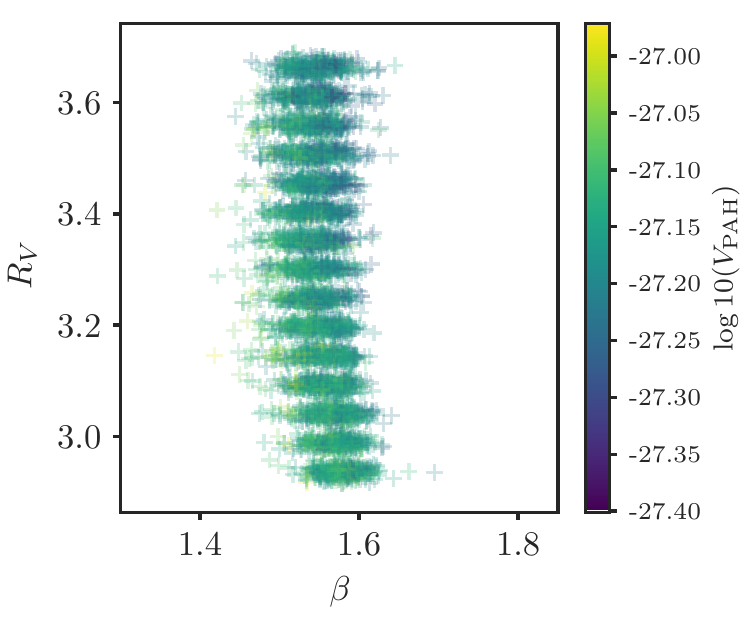} \\
		\vspace{-2.00mm}
		\textbf{(a)}  & \textbf{(b)} & \textbf{(c)}  \\[6pt]
	\end{tabular}
	\caption{$R_{\textnormal{V}}$ vs. $\beta$ for 15 MCMC runs, each corresponding to a distinct Rv value. The points are color coded by the log of the volume of (a) the carbonaceous grains, (b) the silicate grains, and (c) the PAH grains. The volume priors are fixed for carbonaceous grains (graphite+PAH) and silicates to values used in WD01 (\S \ref{sec:methods}). In spite of substantial freedom to explore the space of size distributions, the volume priors plus the constraint to match the S16 reddening curves fail to produce the observed  $R_{\textnormal{V}}$- $\beta$ correlation. This motivates introducing a dependence of the volume priors on $R_{\textnormal{V}}$.
	}
	\label{fig:first_hypothesis_volume_Rv_vs_beta}
\end{figure*}


\begin{figure*}[t!]
	\centering
	\begin{tabular}{cccc}
		\hspace{-5.00mm}
		\includegraphics[width=0.35\textwidth]{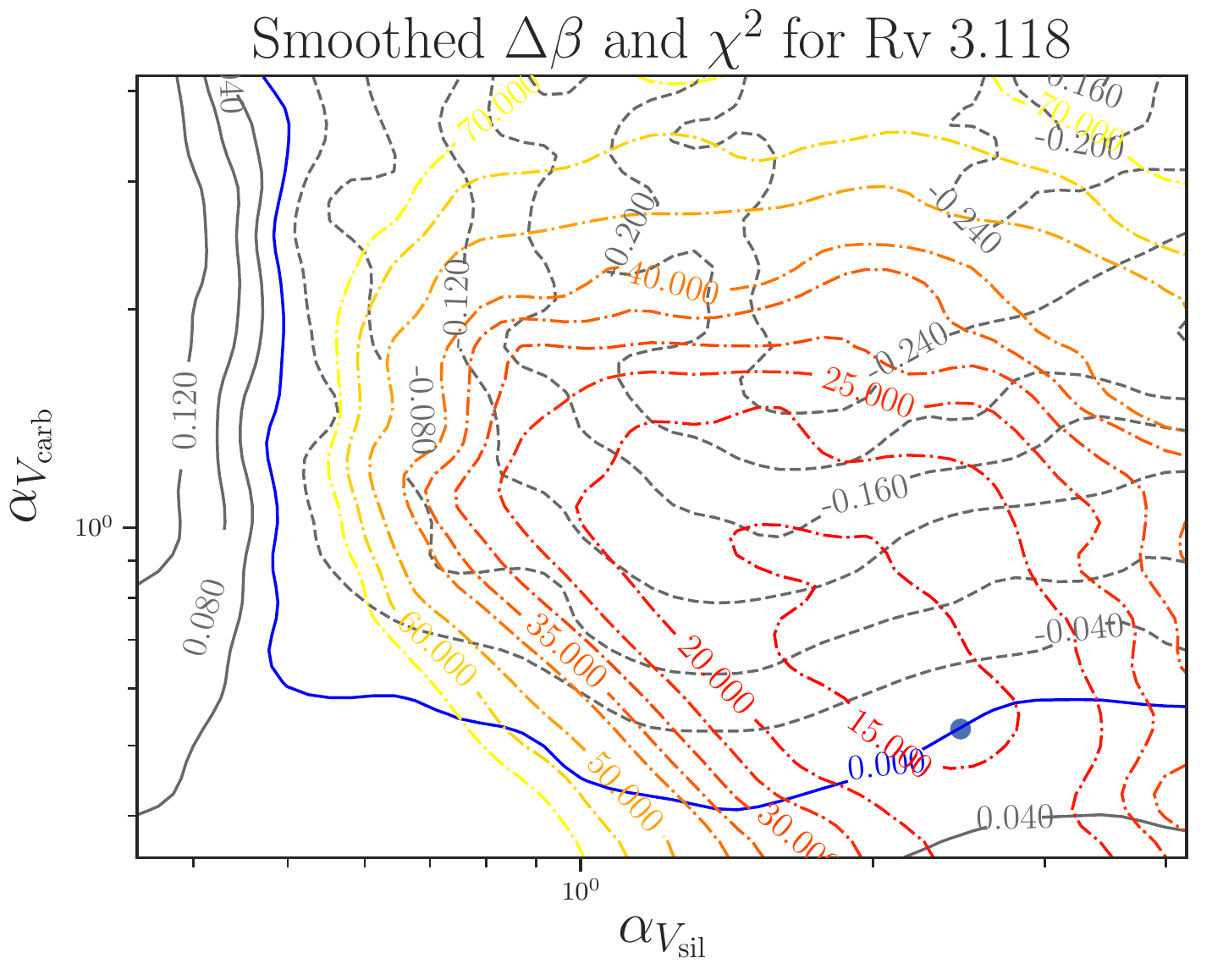} &
		\hspace{-5.00mm}	
		\includegraphics[width=0.35\textwidth]{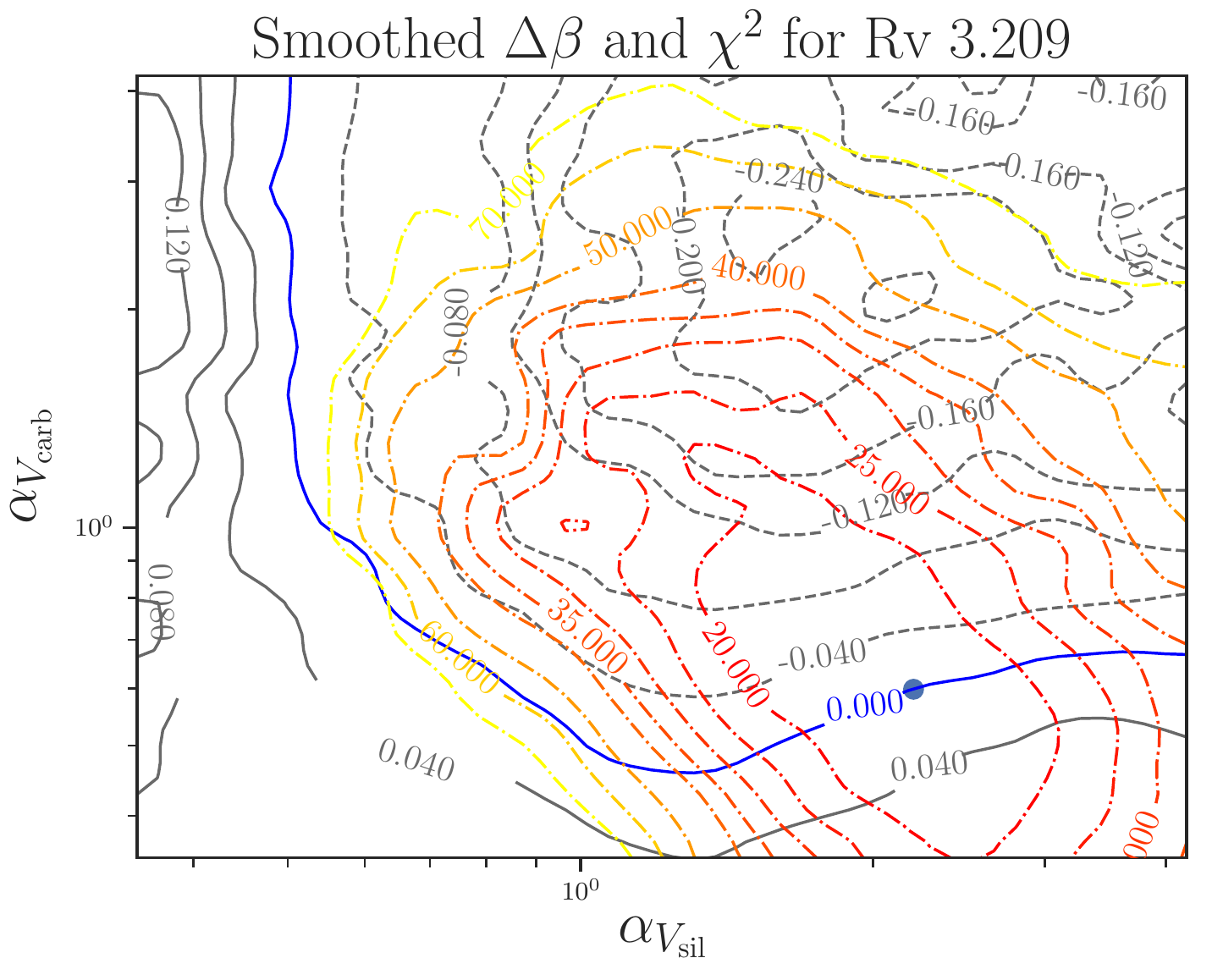} &
		\hspace{-5.00mm}
		\includegraphics[width=0.35\textwidth]{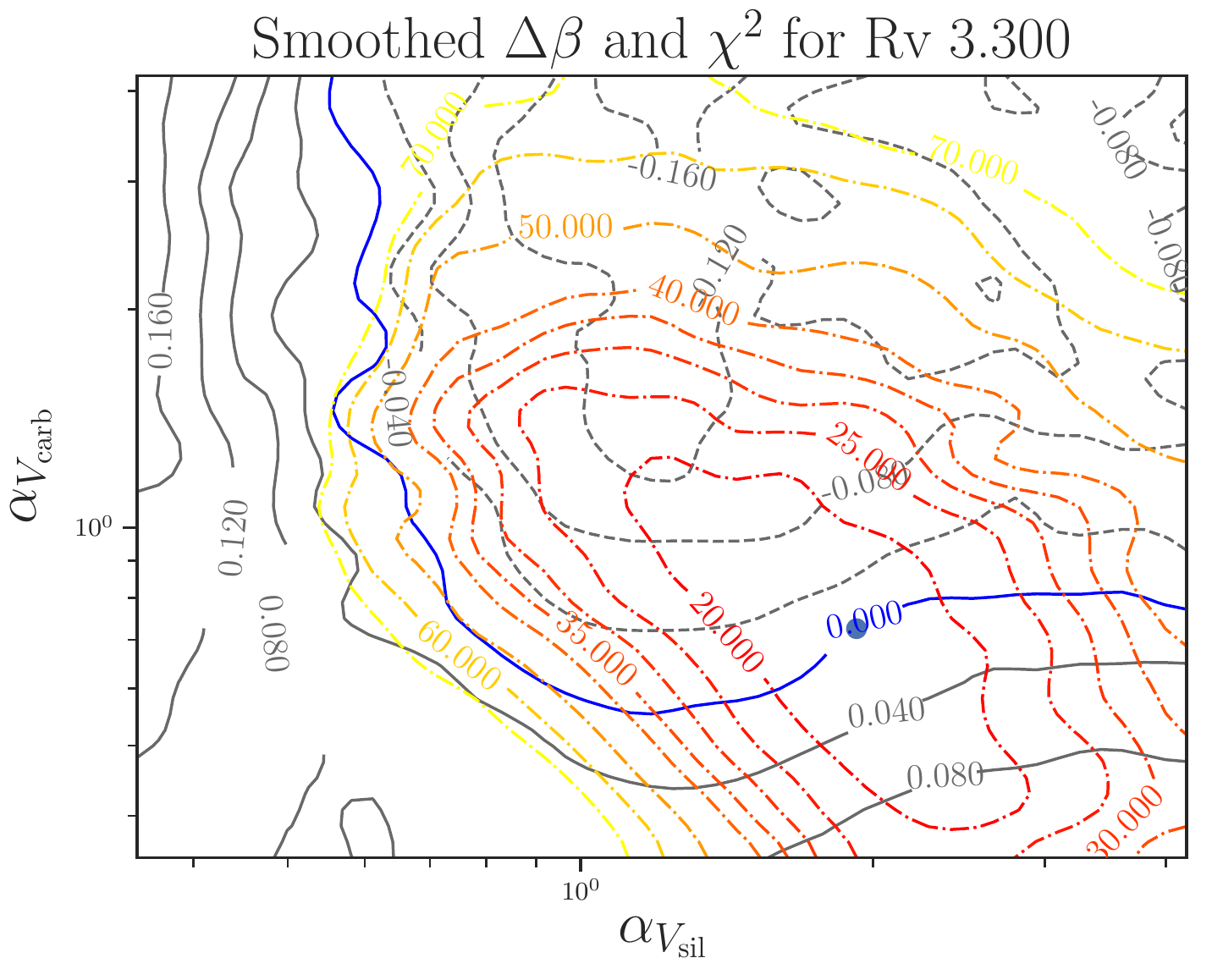} \\
		\vspace{-2.00mm}
		
	\end{tabular}
	
	\vspace{-5.00mm}
	
	\begin{tabular}{cccc}
		\hspace{-5.00mm}
		\includegraphics[width=0.35\textwidth]{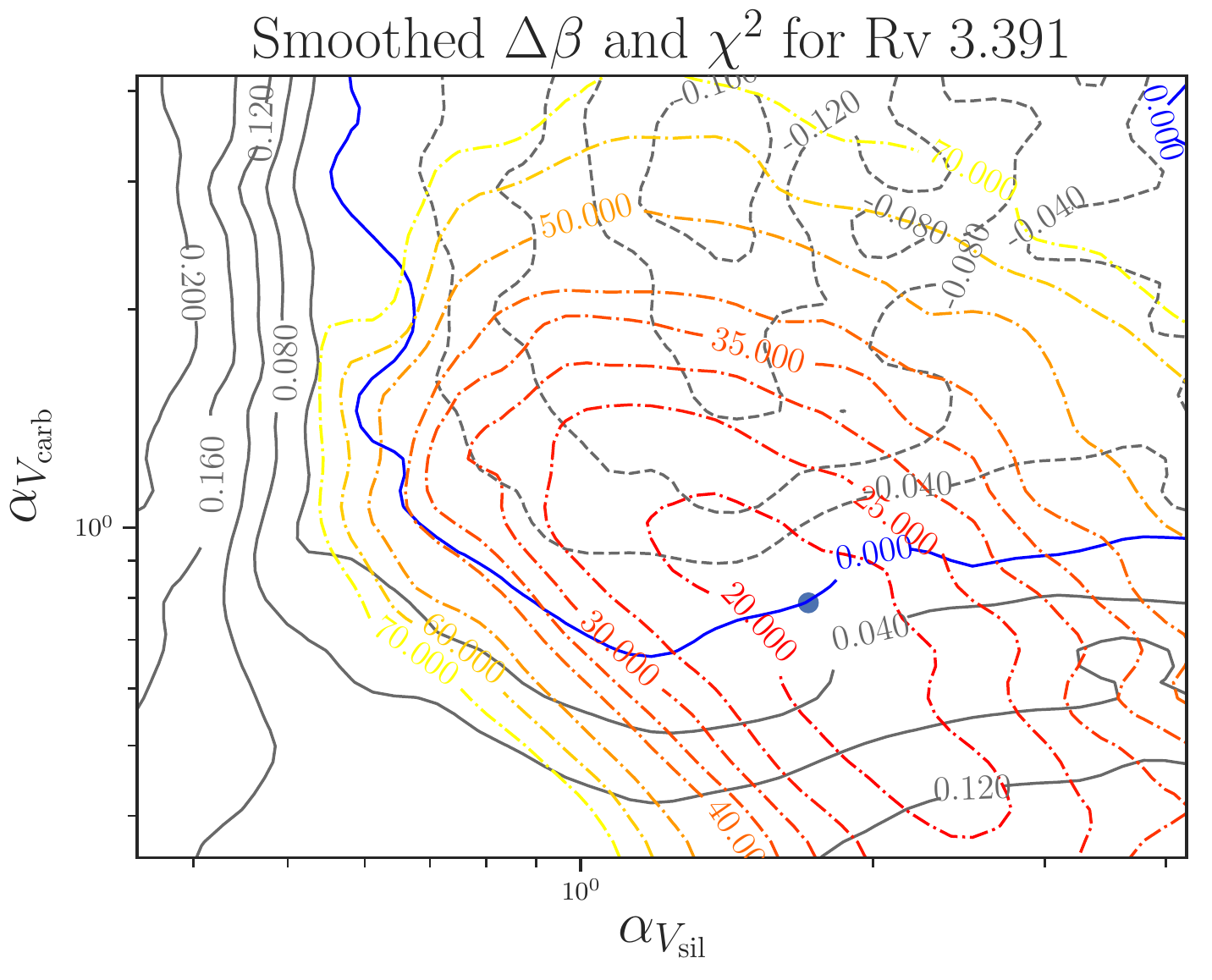} &
		\hspace{-5.00mm}		
		\includegraphics[width=0.35\textwidth]{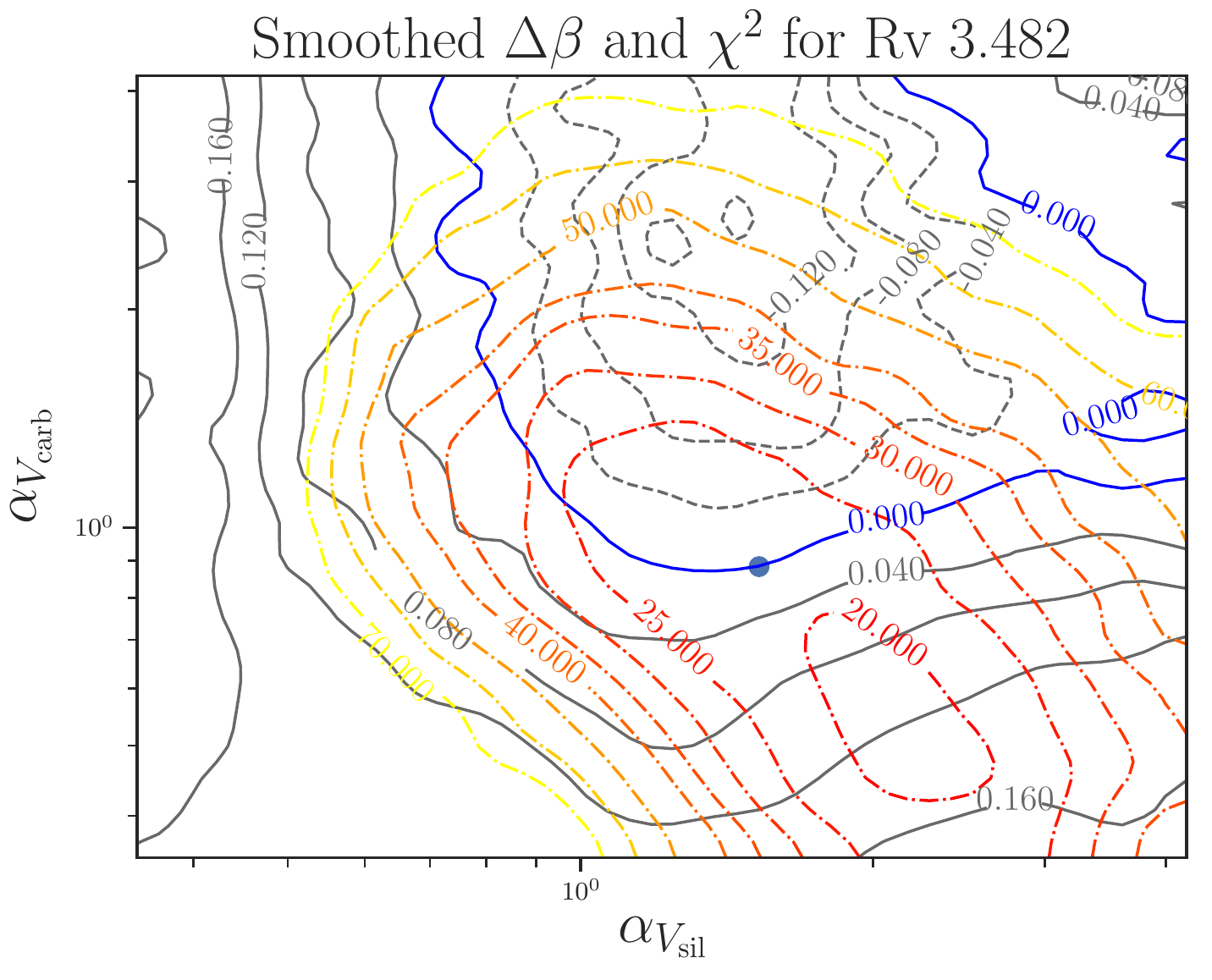} &
		\hspace{-5.00mm}		
		\includegraphics[width=0.35\textwidth]{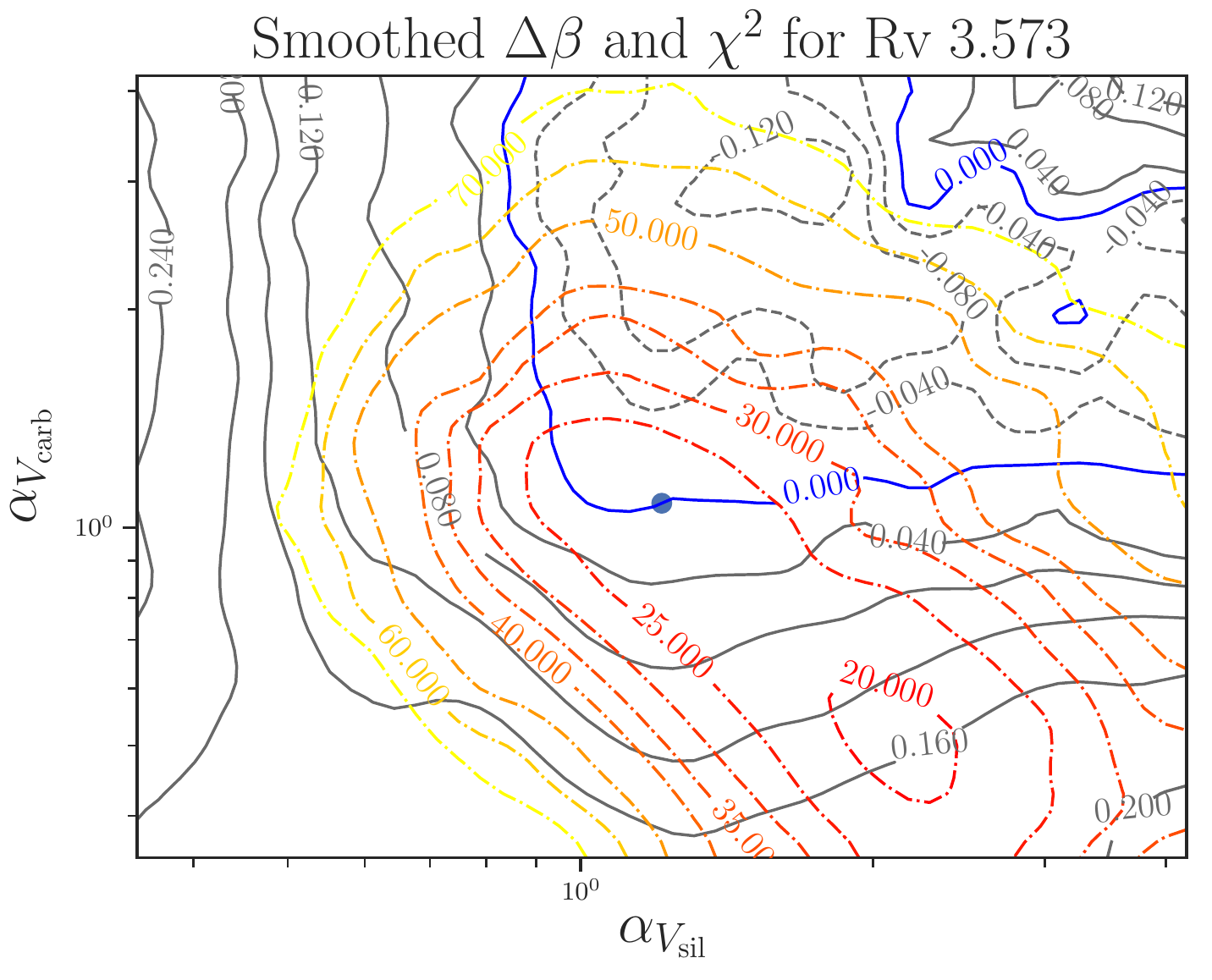} 
		
	\end{tabular}
	\caption{
		Minimum $\chi^2$ (red-orange-yellow dashed-dotted contours ) and $\Delta\beta$ (gray and blue continuous/dashed contours) for 6 values of $R_{\textnormal{V}}$, as a function of carbonaceous and silicate volume coefficients.  The $\Delta\beta$ = 0 contour (blue) corresponds to the S16 empirical  $R_{\textnormal{V}}$-$\beta$ correlation (Eq. \ref{eq:beta_eddie}).  
		In each panel the minimum $\chi^2$ for $\Delta\beta=0$ is marked by a blue dot. The location of these dots shows a monotonic trend as a function of $R_{\textnormal{V}}$.}
	
	\label{fig:optimizer_grid}
\end{figure*}
\paragraph{Likelihood} \label{subsec:likelihood}

We want to compare to Schlafly's reddening vector.
For each proposed set of 11 parameter the MCMC makes, the volume parameters are transformed into the size distribution parameters (converting from a set of $b_\textnormal{C},\ \alpha_g,\  \beta_g,\  a_{t,g},\  a_{c,g},\  V_g,\  \alpha_s,\  \beta_s,\  a_{t,s},\  a_{c,s},\  V_s$ to a set of $b_\textnormal{C},\ \alpha_g,\  \beta_g,\  a_{t,g},\  a_{c,g},\  C_g,\  \alpha_s,\  \beta_s,\  a_{t,s},\  a_{c,s},\  C_s$).
Using the size distribution of the grains, the resulting extinction vector $\bm{A}/N_{\textnormal{H}}$ is calculated at the nine wavelengths (after H was removed when fixing the grey component) from S16 using equation \ref{eq:11_parameters_extinction_equation}. 

Appendix \ref{sec:appendix_error_in_extinction} shows the calculation for the error in extinction that gives the covariance matrix ($\Sigma^{\bm{A''}}$, Eq. \ref{eq:A''_cov}) for each value of $x$, based on the errors in the reddening vectors obtained by S16.

The extinction vector $\bm{A}/N_{\textnormal{H}}$ is compared to the reference extinction vector obtained with Equation \ref{eq:reference_extinction}:

\begin{equation}
\frac{\bm{A}_{\textnormal{residual}}}{N_{\textnormal{H}}} =  \frac{\bm{A}}{N_{\textnormal{H}}} - \frac{\bm{A}_{\textnormal{reference}}}{N_{\textnormal{H}}}
\end{equation}

The likelihood function is $\ln{\mathcal{L}} = -\frac{1}{2} \Delta \chi^2$, with the $\chi^2$ given by:

\begin{equation}\label{eq:likelihood}
\begin{split}
\Delta \chi^2 = \frac{\bm{A}^T_{\textnormal{residual}}}{N_{\textnormal{H}}}(\Sigma^{\bm{A''}})^{-1}  \frac{\bm{A}_{\textnormal{residual}}}{N_{\textnormal{H}}}
\end{split}
\end{equation}
To the likelihood, we add the Gaussian prior on the volume: 
\begin{equation}
\ln{\textnormal{prior}} = -\frac{1}{2}\left(\frac{V_g-V_{g,\textnormal{reference}}}{0.1\cdot V_{g,\textnormal{reference}}}\right)^2-\frac{1}{2}\left(\frac{V_s-V_{s,\textnormal{reference}}}{0.1\cdot V_{s,\textnormal{reference}}}\right)^2
\end{equation}
to obtain the posterior:
\begin{equation}
\ln{\textnormal{posterior}} = \ln{\textnormal{prior}} + \ln{\mathcal{L}}
\label{eq:ln_posterior}
\end{equation}
Here $V_g$ represents the sum of the volumes for the PAH and carbonaceous grains, and $V_s$ the silicate grains.

\subsection{Exploring the Dust Parameters' Posterior Distribution with an MCMC}\label{sec:method_MCMC}

We sample from the posterior (Eq. \ref{eq:ln_posterior}) for a target extinction curve at fixed  $R_{\textnormal{V}}$  (\S \ref{sec:extinction_modeling}) for each of 15 values of $R_{\textnormal{V}}$ linearly spaced between 2.94 and 3.67. The posterior is conditional on $R_{\textnormal{V}}$ instead of letting $R_{\textnormal{V}}$ float, so that the uncertainty in the target extinction curve does not depend on the parameters at each step in the Markov chain.
To expedite burn-in, we initiate the MCMC at a set of dust grain size distribution parameters determined by optimization.

The MCMC uses the \texttt{ptemcee} \footnote{The code can be found at the Python repository at \url{https://pypi.org/project/ptemcee/} or at Will Vousden github repository at \url{https://github.com/willvousden/ptemcee}} \cite{Vousden2016} package that uses parallel tempering. This allows for a much more efficient exploration of the parameter space than something like the Metropolis-Hastings algorithm. We experimented with a number of temperatures between 3 and 5, and  found that there was no significant difference in the results, so we settled for 3 temperatures to reduce the computational time. 300 walkers were run for each $R_{\textnormal{V}}$, for 100,000 steps.

\subsection{Studying the Correlation between Dust Emissivity and Absorption}\label{sec:beta_Rv_correlation} 

Taking the final posterior distributions from all of the chains, and using the precomputed values of the temperature $T$ for each radius of the grain (Fig. \ref{fig:equilibrium_temperature}),  we integrate to calculate the specific intensity, at each of the four bandpass frequencies of Planck.

The emitted radiation is modeled as a modified black body shown using  Eq. \ref{eq:MBB}, and fit to find the three parameters $ (\tau_{353} ,\beta, T)$ corresponding to each sample from the MCMC. We generate the emission at the four wavelength bands corresponding to the Planck satellite, and then fit the four data points to the modified black body law, using corresponding weighting and bandpass filters as used by the Planck team.

Next, we  calculate the $R_{\textnormal{V}}$ for the sample, and see if there is any correlation between $\beta $ and $R_{	\textnormal{V}}$, thus comparing to the results in S16 (Fig. \ref{fig:Rv_vs_beta}).


\section{Results and Discussion}\label{sec:results_and_discussion}

\subsection{Correlation between dust extinction and far-infrared emissivity}

We consider 2 hypotheses for the origin of the $R_{\textnormal{V}}-\beta$ correlation:
\begin{enumerate}[label=\Roman*.]
	\item the size distribution hypothesis, which attributes the variation in $R_{\textnormal{V}}$ and $\beta$ to variations in grain size distribution, holding the volume (per H) of each species fixed\footnote{However, the Kramers-Kronig relation can be used to determine a lower bound on the volume of the grains for a given extinction function integrated over a finite wavelength interval \citep{Purcell1969}. \cite{Mishra2017} applied this relation to approximate the volumes for the silicate and carbonaceous grains. To perform this calculation, we would have to integrate over the UV part of the absorption, which is a bit uncertain \citep{Peek2013}, so it was not included in this work. Nevertheless, one should keep in mind that the bounds that we are proposing might be violating this relation slightly.}. 

	\item the composition hypothesis, which requires variation in the relative volumes of silicates and carbonaceous grains to vary as a function of $R_{\textnormal{V}}$.
\end{enumerate}

Fixing the volume priors in hypothesis I proved to be too restrictive: the MCMC does not explore the full range of the $\beta$ parameters (Fig. \ref{fig:first_hypothesis_volume_Rv_vs_beta}).  This fails to yield the observed $R_{\textnormal{V}}-\beta$ correlation.

If an $R_{\textnormal{V}}$-dependent prior on volumes is needed for hypothesis II, what form should it take?

\subsubsection{Optimizer Results}\label{sec:optimizer_results}

To generate a hypothesis about what form of volume priors might give rise to the $R_{\textnormal{V}}-\beta$ correlation, we begin by mapping out the volume parameter space $V_s$, $V_g$. 

We use an optimizer to explore the effects of the volume priors on the resulting $\beta$. This framework is used instead of the MCMC in the beginning of the analysis in order to take advantage of the great increase in speed, which allows us to explore the parameter space of volume priors on a much finer grid. Thus, we can get an idea fast of what parameter combinations would provide useful results.

The Gaussian volume priors are defined to be centered at $V_{\textnormal{sil}} = \alpha_{V_{\textnormal{sil}}}\cdot V_{0,V_{\textnormal{sil}}}$, $V_{\textnormal{carb}} = \alpha_{V_{\textnormal{carb}}}\cdot V_{0,V_{\textnormal{carb}}}$ where $\alpha_{V_{\textnormal{sil}}}$ and $\alpha_{V_{\textnormal{carb}}}$ are control parameters that we use to scale the total carbonaceous and silicate volumes up and down. $V_{0,V_{\textnormal{sil}}}$ and $V_{0,V_{\textnormal{carb}}}$ are the reference fiducial values for the volumes from WD01, as described in section \ref{sec:methods}. We let $\alpha_{V_{\textnormal{sil}}}$ and $\alpha_{V_{\textnormal{carb}}}$ take values between 0.35 and 4.2, and sample the interval logarithmically at 50 values, creating a 50 by 50 grid for each $R_{\textnormal{V}}$.

For each of the points in the grid, the $\chi^2$ returned by the optimizer is calculated. We smooth over the resulting image using a Gaussian filter with $\sigma = 1.5$, and calculate the contour plots over the resulting image.

For each $R_{\textnormal{V}}$ panel, to compare with the expected $\beta$, we use a best fit line obtained from the $R_V $ vs. $\beta$ data set from S16, given by eq \ref{eq:beta_eddie}.
\begin{equation}
\beta_{\textnormal{Schlafly}}(R_{\textnormal{V}}) = 2.82-0.36\cdot R_{\textnormal{V}}
\label{eq:beta_eddie} 
\end{equation}
Thus, for each $\beta$ obtained from the points in the grid, we calculate $\Delta \beta = \beta-\beta_{\textnormal{Schlafly}}$. Using a Gaussian filter with $\sigma = 1.5$, we smooth over the resulting image, and calculate the contour plots. The resulting contour plots for both the $\chi^2$ and $\Delta \beta$ analysis are superimposed (Fig. \ref{fig:optimizer_grid}). Each grid corresponds to a different $R_{\textnormal{V}}$ value.
\begin{figure}[t]
	\includegraphics[scale=1]{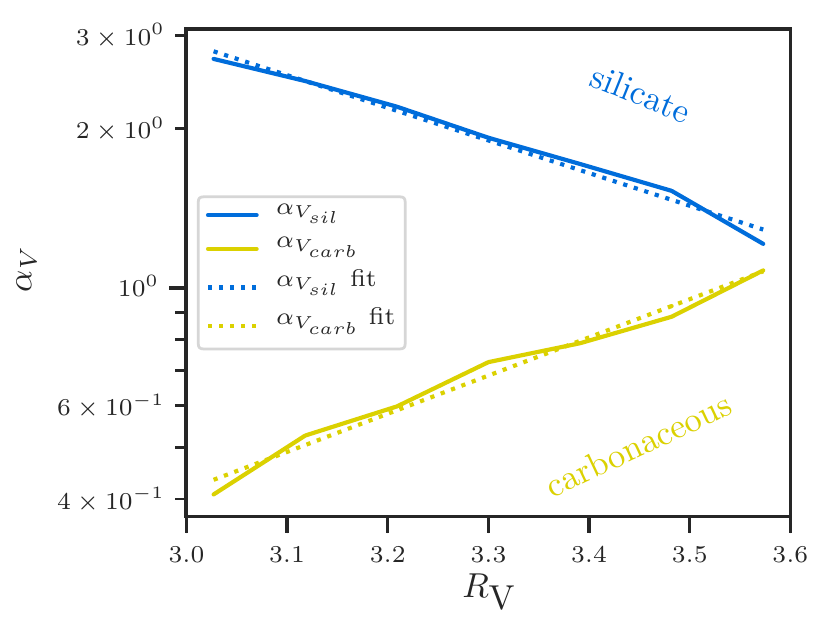}
	\caption{$\alpha_V$ is the ratio of the volume relative to Draine prior. Using the data from the optimizer, we fit $\ln{\alpha_V}$ as linear functions of $R_{\textnormal{V}}$, obtaining a decreasing function for silicates and an increasing function for carbonaceous grains.}
	\label{fig:optimizer_function_fit}
\end{figure}

For each grid, we find the point of minimum $\chi^2$ that has $\Delta \beta =0$. We read the corresponding $\alpha_{V_{\textnormal{carb}}}$ and $\alpha_{V_{\textnormal{sil}}}$ values, and plot them against $R_{\textnormal{V}}$ (Fig. \ref{fig:optimizer_function_fit}). The points show a log-linear dependence. Performing  linear fits of $\ln{\alpha_{V_{\textnormal{carb}}}}$ vs. $R_{\textnormal{V}}$ and $\ln{\alpha_{V_{\textnormal{sil}}}}$ vs. $R_{\textnormal{V}}$, the functions shown in equations \ref{eq:volume_sil_fit_function} and \ref{eq:volume_carb_fit_function} are obtained.

\begin{equation}
\ln{\alpha_{V_{\textnormal{sil}}}}(R_{\textnormal{V}}) = -1.42\cdot R_{\textnormal{V}}+5.32
\label{eq:volume_sil_fit_function}
\end{equation}

\begin{equation}
\ln{\alpha_{V_{\textnormal{carb}}}}(R_{\textnormal{V}}) = 1.66\cdot R_{\textnormal{V}}-5.85
\label{eq:volume_carb_fit_function}
\end{equation}

This relation of the volume priors on $R_{\textnormal{V}}$ also depends on extinction per $N(H)$, assumed to be $A_{\textnormal{I}}/N_{\textnormal{H}}= 3.38\times10^{-22}\textnormal{cm}^2$ (beginning of \S \ref{sec:methods}).  The $R_{\textnormal{V}}$-$\beta$ relation itself does not depend on this assumption, because neither $R_{\textnormal{V}}$ nor $\beta$ depend on the column density per se. However an increase in the assumed  $A_{\textnormal{I}}/N_{\textnormal{H}}$ would require an increase in the volume (per H) of each species, by the same factor. In other words, a different  $A_{\textnormal{I}}/N_{\textnormal{H}}$ convention would simply slide the contours in Fig. \ref{fig:optimizer_grid} up and to the right. If a different $A_{\textnormal{I}}/N_{\textnormal{H}}$ is chosen, such as $A_{\textnormal{I}}/N_{\textnormal{H}} =2.6\times10^{-22}\textnormal{cm}^2$ \citep{Zhu2017}, the volume results can in turn be scaled by 0.77 = 2.6/3.38 and obtain the same behavior.

\setlength{\tabcolsep}{4pt} 
\begin{table*}[t]
\centering
\begin{tabular}{lllllllllllll}
\toprule\toprule
  $x$& $R_V$   & $10^5$ bC & $\alpha_g$ & $\beta_g$ & $a_{t,g}[\mu $m] & \
                $a_{c,g}[\mu $m] & $10^{27}V_g [\textnormal{cm}^3$ $\textnormal{H}^{-1}]$   \
                & $\alpha_s$ & $\beta_s$ & $a_{t,s}[\mu $m] & $a_{c,s}[\mu $m] & $10^{27}V_s [\textnormal{cm}^3$ $\textnormal{H}^{-1}]$   
 \\ \midrule 
-0.040&2.94&3.00&-0.66&0.07&0.00035&0.68&0.813&-1.83&-5.41&0.16&0.19&8.355
  \\ \midrule
-0.034&2.99&3.00&-0.72&0.12&0.06709&0.73&0.871&-1.60&-23.45&0.13&0.21&8.385
  \\ \midrule
-0.029&3.04&3.00&-0.83&-0.00&0.00035&1.02&0.975&-1.59&-20.81&0.22&0.11&7.460
  \\ \midrule
-0.023&3.09&3.00&-1.47&-0.00&0.00282&1.27&1.047&-1.56&-19.47&0.17&0.16&7.364
  \\ \midrule
-0.017&3.14&3.00&-1.05&-0.17&0.00883&1.34&1.111&-1.94&-2.78&0.21&0.14&7.187
  \\ \midrule
-0.011&3.20&3.00&-1.49&-0.00&0.00035&1.72&1.204&-1.65&-5.28&0.16&0.18&6.662
  \\ \midrule
-0.006&3.25&4.70&-1.30&-0.01&0.00119&1.41&1.315&-1.76&-8.79&0.35&0.00&6.496
  \\ \midrule
0.000&3.30&4.66&-1.30&-3.08&0.00036&1.54&1.466&-1.78&-2.72&0.26&0.09&5.683
  \\ \midrule
0.006&3.35&4.11&-2.50&0.00&0.00141&1.53&1.526&-1.23&-6.84&0.14&0.20&5.176
  \\ \midrule
0.011&3.40&5.40&-1.93&-0.01&0.00241&2.58&1.672&-1.05&-30.00&0.16&0.17&4.944
  \\ \midrule
0.017&3.46&5.81&-2.56&0.00&0.00035&1.78&1.868&-0.80&-20.84&0.18&0.14&4.176
  \\ \midrule
0.023&3.51&5.94&-2.01&-0.00&0.00065&4.83&1.991&-0.52&-19.04&0.09&0.22&4.126
  \\ \midrule
0.029&3.56&6.09&-1.67&-0.53&0.00041&2.48&2.110&-0.61&-4.43&0.15&0.17&3.684
  \\ \midrule
0.034&3.61&6.07&-2.85&0.02&0.01434&1.85&2.303&-0.87&-10.25&0.33&0.02&3.654
  \\ \midrule
0.040&3.66&6.41&-1.78&-0.25&0.00047&3.68&2.406&-0.50&-3.92&0.19&0.14&3.325
  \\ \midrule
 \bottomrule
\end{tabular}
\caption{Optimized parameter values of the dust grain size distributions for each $x$/$R_V'$, for hypothesis II.\label{table:optimized_dust_table}}
\end{table*}

\subsubsection{MCMC Analysis}\label{sec:MCMC_analysis_results}
\begin{figure}[t]
	\includegraphics[scale=1]{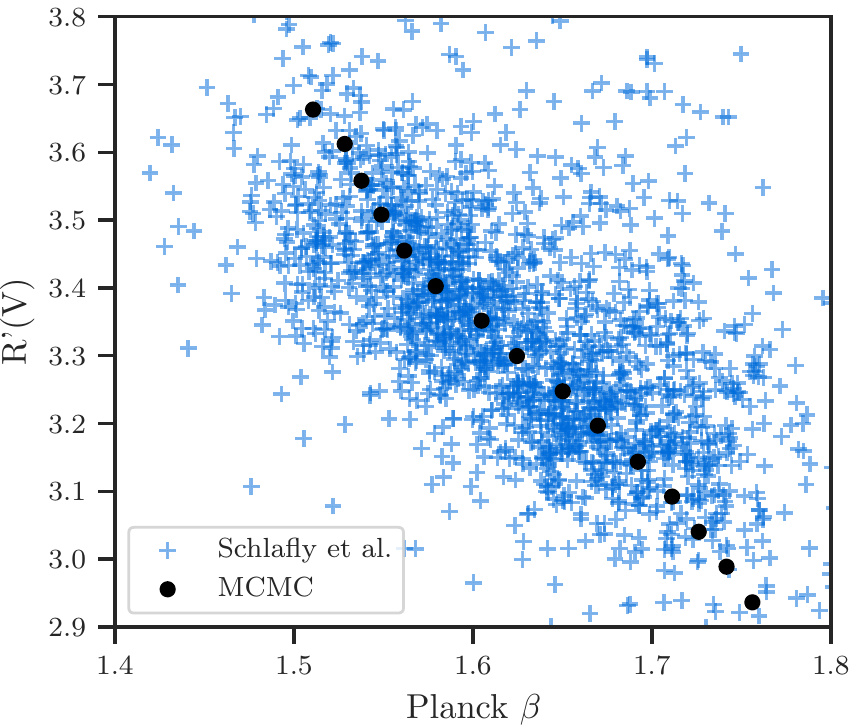}
	\caption{An MCMC analysis if performed for 15 values of $R_{\textnormal{V}}$, using dedicated volume priors as a function of $R_{\textnormal{V}}$ (\S \ref{sec:optimizer_results}). For each of the 15 posterior clouds, the average $R_{\textnormal{V}}$ and $\beta$ are obtained, seen here superimposed on the data from S16. The anti-correlation relation trend between $R_{\textnormal{V}}$ and $\beta$ is thus reproduced. In this work $R_{\textnormal{V}}$ refers to the $R_{\textnormal{V}}'$ from S16.}
	\label{fig:Rv_vs_beta}
\end{figure}

\begin{figure}[t]
	\includegraphics[scale=1]{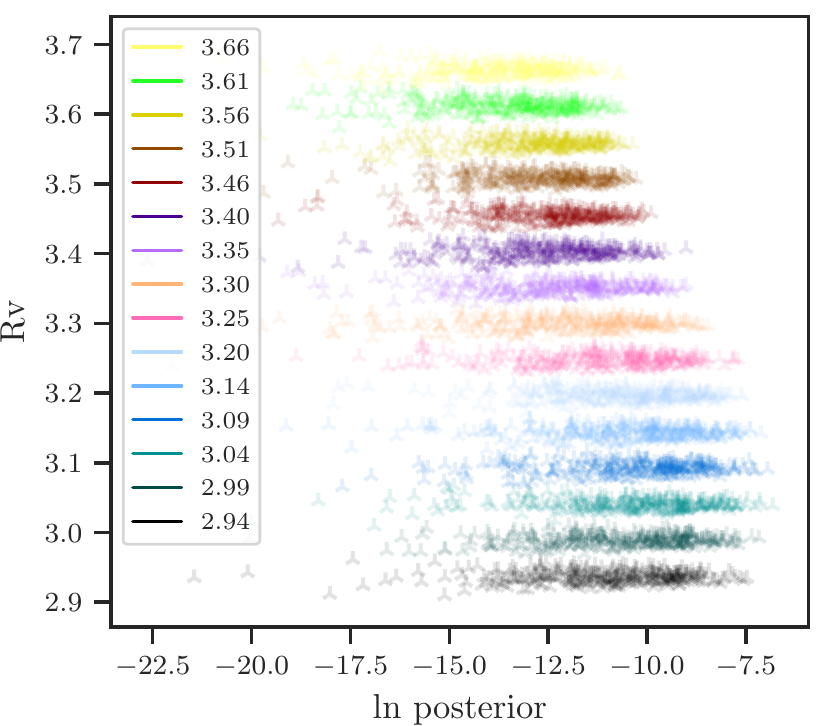}
	\caption{Log of the posterior for each point in the 15 runs of the MCMC corresponding to a distinct $R_{\textnormal{V}}$ value. While there is variation in the posterior values, this variation is contained in the range between ln posterior of -7.5 and -20. The modest range in $\ln{P}$ is reassuring.}
	\label{fig:Rv_vs_posterior}
\end{figure}

\begin{figure*}[t]
	\centering
	\begin{tabular}{cccc}
		\hspace{-5.00mm}
		\includegraphics[width=0.35\textwidth]{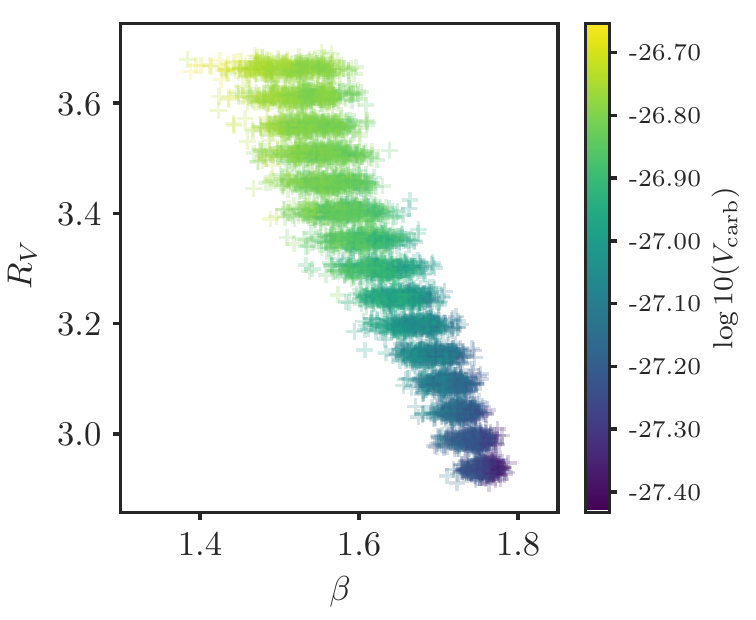} &
		\hspace{-5.00mm}	
		\includegraphics[width=0.35\textwidth]{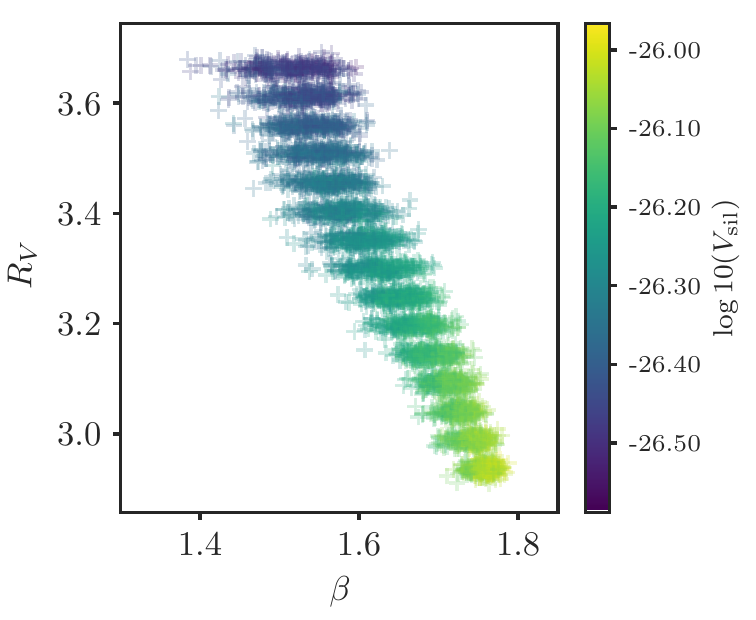} &
		\hspace{-5.00mm}
		\includegraphics[width=0.35\textwidth]{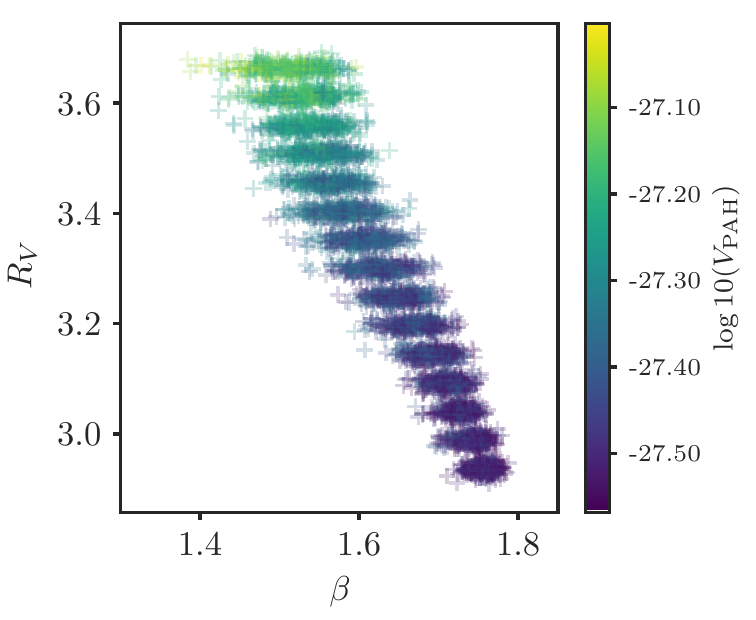} \\
		\vspace{-2.00mm}
		\textbf{(a)}  & \textbf{(b)} & \textbf{(c)}  \\[6pt]
	\end{tabular}
	\begin{tabular}{cccc}
		\hspace{-5.00mm}
		\includegraphics[width=0.35\textwidth]{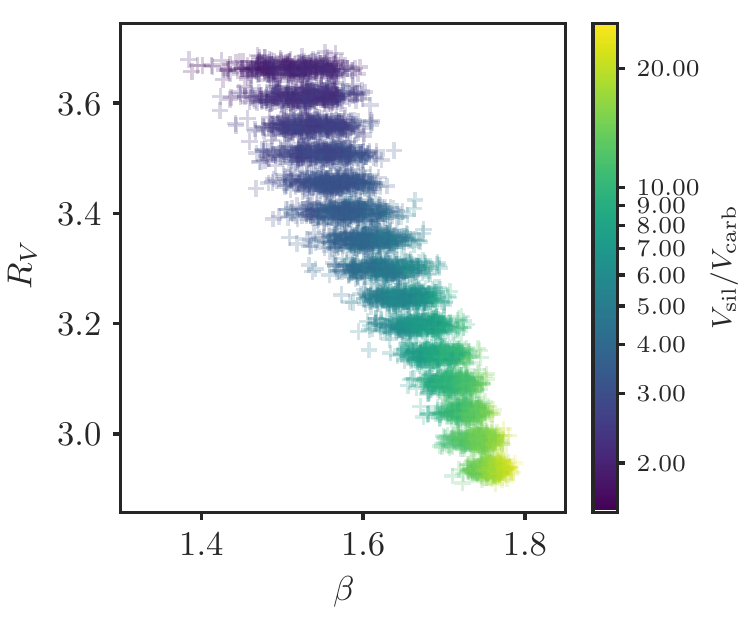} &
		\hspace{-5.00mm}		
		\includegraphics[width=0.35\textwidth]{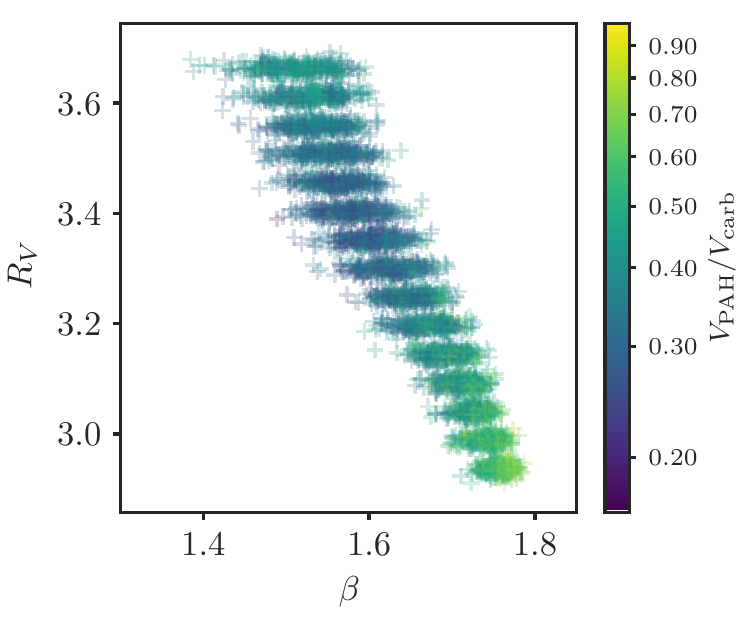} \\
		\textbf{(d)}  & \textbf{(e)}  \\[6pt]
	\end{tabular}
	\caption{The points in the MCMC cloud are here color coded by \textbf{(a)} the $\log10$ of the volume of the carbonaceous grains, 
		\textbf{(b)}  the $\log10$ of the volume of the silicate grains, 
		\textbf{(c)}  the $\log10$ of the volume of the PAH grains, 
		\textbf{(d)} the ratio of the volumes of silicate grains and the volume of carbonaceous grains, and
		\textbf{(e)} the ratio of the volumes of PAH grains and the volume of carbonaceous grains.
	We find that a composition with higher ratio of carbonaceous to silicate grains leads to more $R_{\textnormal{V}}$ and lower $\beta$. Carbonaceous grains are the sum of PAH and graphite grains. PAH and graphite also increase with $R_{\textnormal{V}}$ independently.}
	\label{fig:volume_Rv_vs_beta}
\end{figure*}

The fits described in Section \ref{sec:optimizer_results} represent a hypothesis for how carbonaceous and silicates volumes might change as a function of $R_{\textnormal{V}}$, and now we must validate that hypothesis using an MCMC analysis.

The resulting values for each parameter can be seen in Table \ref{table:optimized_dust_table}.
Using the functions found in equations \ref{eq:volume_sil_fit_function} and \ref{eq:volume_carb_fit_function}, we now set up volume priors correspondingly, and turn to performing an MCMC analysis as described in the \S \ref{sec:method_MCMC}. The optimizer is run with the new volume priors. The resulting values for each parameter (Table \ref{table:optimized_dust_table}) are used as initializing points for the MCMC. We run 15 MCMCs for 15 $ R_{\textnormal{V}}$ values linearly spaced between 2.936 and 3.664 ($x$ varies between -0.04 and 0.04). For each of the MCMC runs, we calculated the dust emissivity and the $R_{\textnormal{V}}$ for each sample from the posterior, using the procedure described in \S \ref{sec:beta_Rv_correlation}

The results can be seen in Fig. \ref{fig:Rv_vs_beta}. The spectral index $\beta$ is spread between 1.4 and 1.8. The variation in spectral index value is significant, which indicates that having different size distributions of dust grains in different directions of the sky can motivate the need to model these different lines of sight with different spectral index values. The values of $R_{	\textnormal{V}}$ and $\beta$ obtained are in the same range as the ones obtained by S16. Most importantly, our values reproduce the trend of the $R_{\textnormal{V}}-\beta$ anticorrelation. The log posterior of all the end positions of the chains in the runs is contained within a modest range (Fig. \ref{fig:Rv_vs_posterior}). The systematic uncertainty in the Planck $\beta$ measurements is thought to be of order 0.05.

\paragraph{Volume and Composition} As expected for hypothesis II, we find that a composition with higher ratio of carbonaceous to silicate grains leads to higher $R_{\textnormal{V}}$ and lower $\beta$ (Fig. \ref{fig:volume_Rv_vs_beta}).

While the functions \ref{eq:volume_sil_fit_function} and  \ref{eq:volume_carb_fit_function} seem very precise, they do not represent unique solutions. We are making a plausibility argument, not a final determination of model parameters for a firmly established model. One might be able to find other solutions that explain the  $R_{\textnormal{V}}-\beta$ correlation. But it is suggestive that $R_{\textnormal{V}}$-dependent volume priors give this behavior, and fixed priors do not.

\begin{figure*}[t!]
	\includegraphics[scale=1]{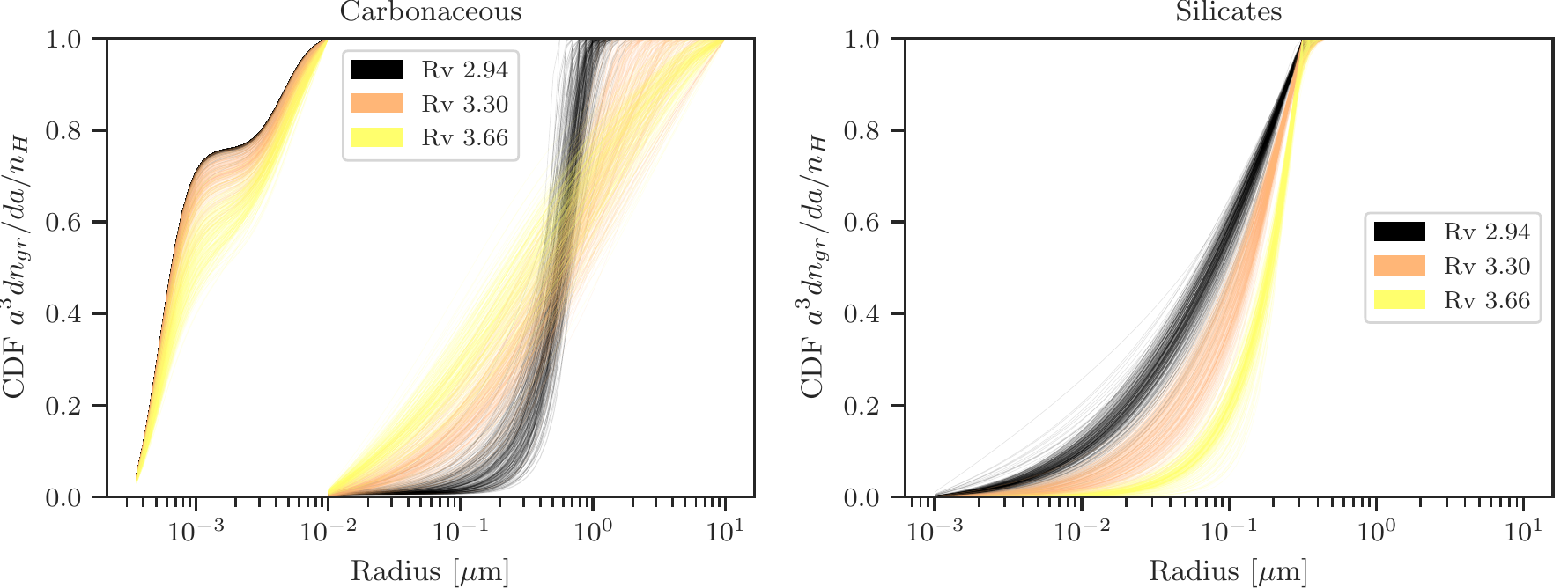}
	\caption{The cumulative distribution function (CDF) corresponding to the volume of the grains. For both silicates and carbonaceous grains, we can see that as one moves to higher $R_{\textnormal{V}}$ , at least 50$\%$ of the volume is in grains of larger and larger size. This is in accordance with the expectation that larger grains lead to higher $R_{\textnormal{V}}$. For the carbonaceous grains, we represented the CDF separately for PAH grains at radii smaller than 0.01$\mu$m and graphite grains at radii larger than 0.01$\mu$m. For the PAHs at a low $R_{\textnormal{V}}$ of 2.94, the size distribution is constrained tightly through the $b_{\textnormal{C}}$ parameter, which results in reduced variation represented by the very thin black line.}
	\label{fig:MCMC_CDF_mass_grains}
\end{figure*}

\begin{figure*}[t!]
	\includegraphics[scale=1]{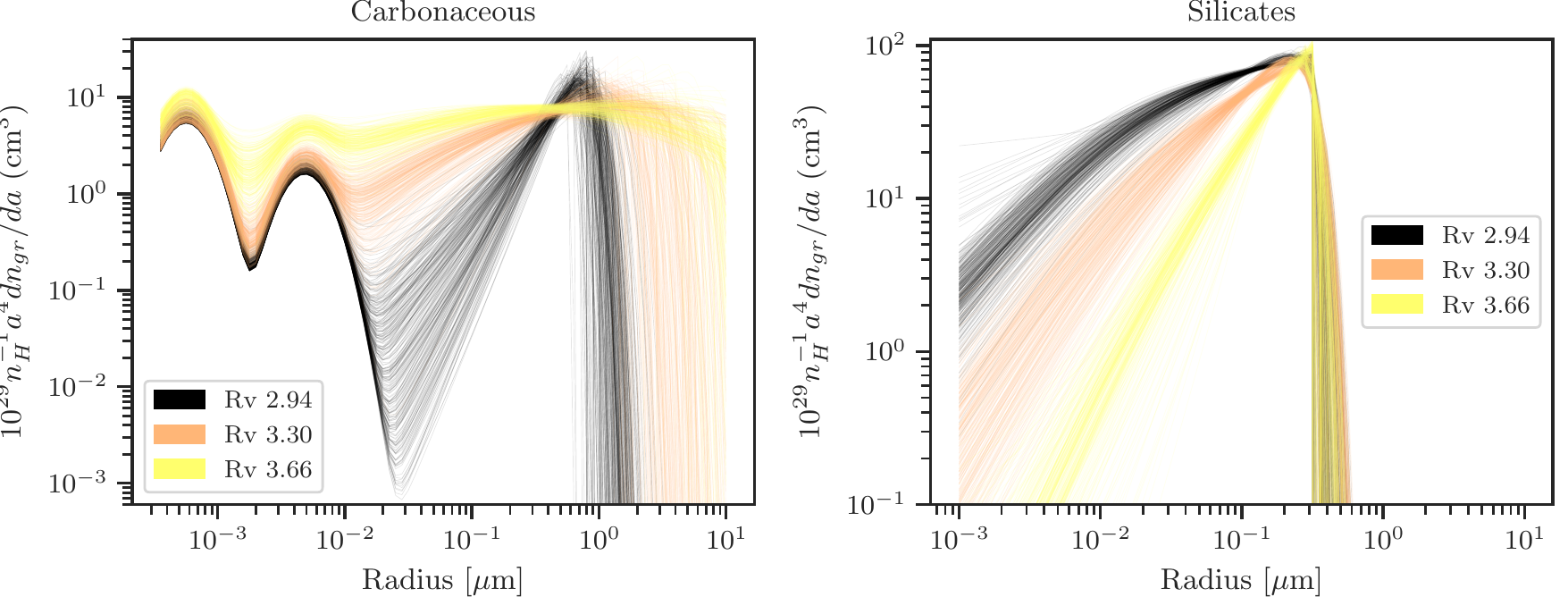}
	\caption{The size distributions corresponding the points in the posterior clouds of from 3 MCMCs at different $R_{\textnormal{V}}$s. The two bumps in the carbonaceous grains at radii smaller than 0.01$\mu$m come from the constraints imposed on the minimum value of the $b_{\textnormal{C}}$ parameter that informs the amount of PAH. We see that larger $R_{\textnormal{V}}$ leads to larger grains cutoffs, but also to a larger ratio of carbonaceous to silicate grains.}
	\label{fig:MCMC_size_dist}
\end{figure*}

\begin{figure*}[t!]
	\includegraphics[scale=1]{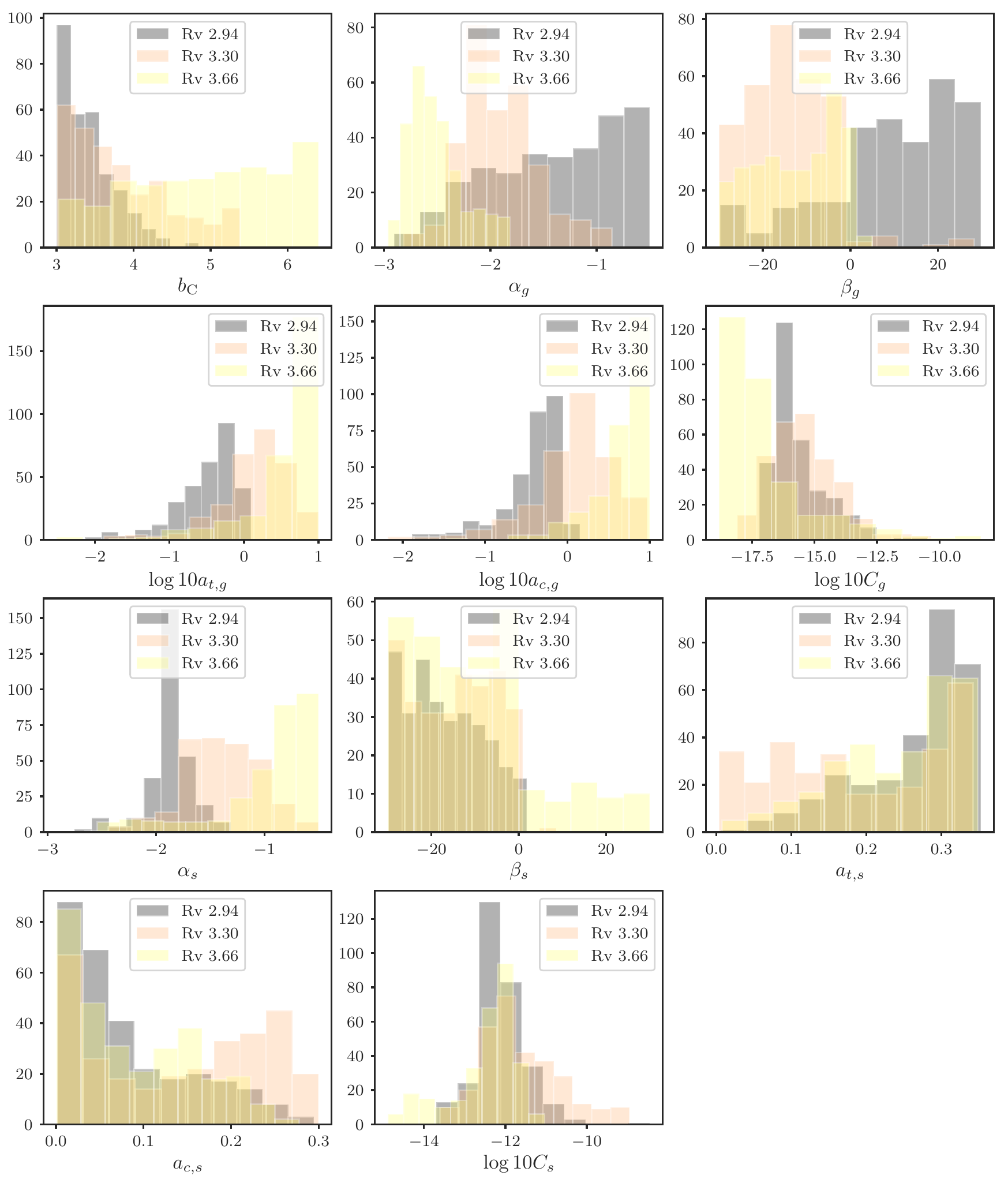}
	\caption{Binned histograms of each of the 11 parameters of the size distributions, shown for 3 different $R_{\textnormal{V}}$ MCMC runs. The distributions for each of the 11 parameters change substantially as $R_{\textnormal{V}}$ changes, and are largely distinct from each other.}
	\label{fig:param_bined_distributions}
\end{figure*}

\begin{figure*}[t]
	\includegraphics[scale=1]{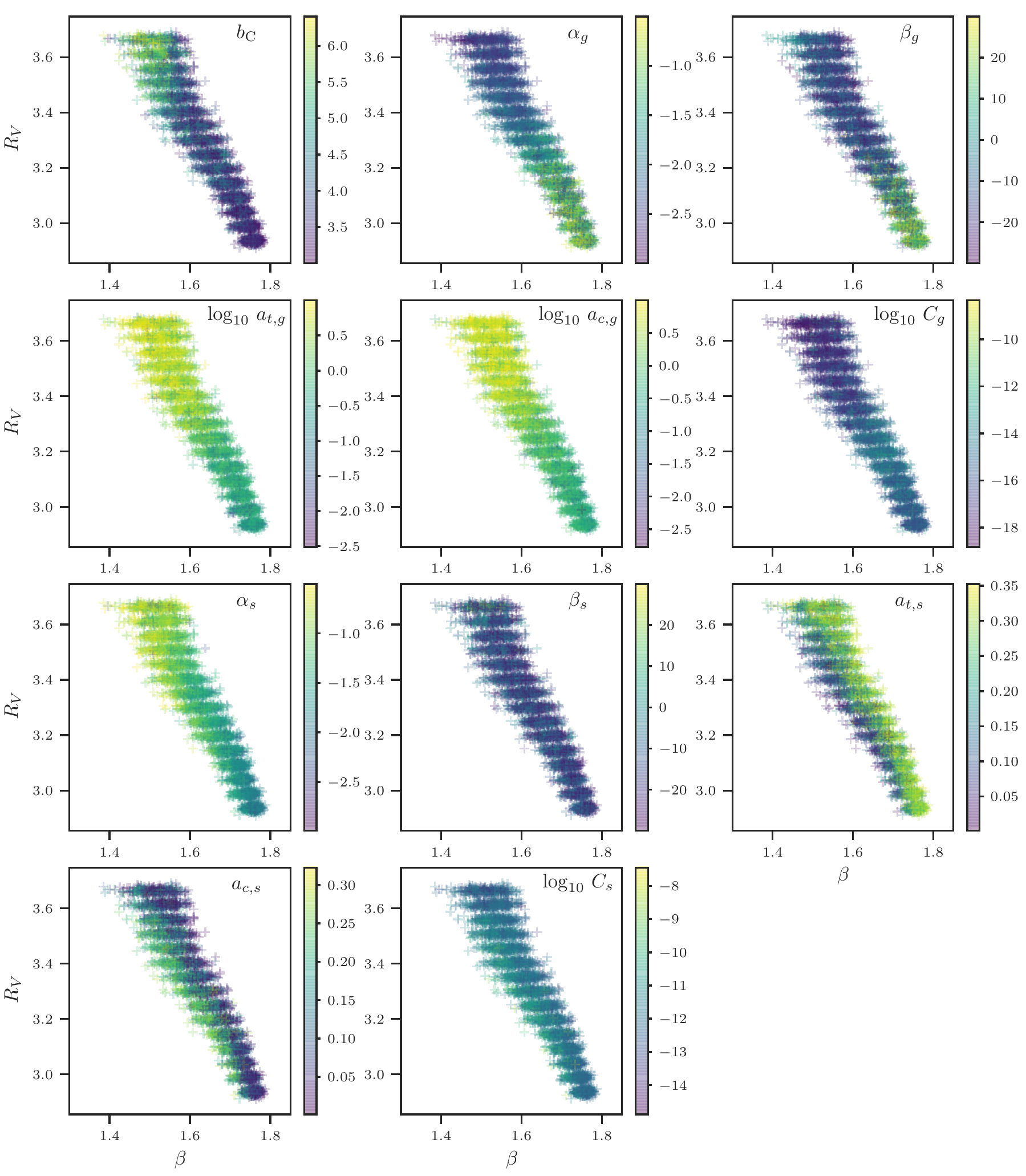}
	\caption{The points from the MCMC clouds colorcoded using the values of the 11 size distribution parameters. $a_{c,s}$ and $a_{t,s}$ are anti-correlated; this anti-correlation can explain the spread of $\beta$ values at fixed $R_{\textnormal{V}}$. If we hold the ratio of carbonaceous to silicates, ratio of PAH to graphite and $R_{\textnormal{V}}$ constant, we can push $\beta$ back and forth by about 0.05 (a relatively small amount) by changing the large grains cutoff. }
	\label{fig:param_Rv_vs_beta}
\end{figure*}

\paragraph{Size of grain distribution} One of the intuitive expectations of the analysis was that larger grains would lead to higher $R_{\textnormal{V}}$. In order to test this hypothesis, we plot the cumulative distribution function of the volume of the grains versus radii. Fig. \ref{fig:MCMC_CDF_mass_grains} shows what percentage of the volume of the grains is made up of radii smaller than each possible radius value. For example, for the case of silicates,  80$\%$ of low $R_{\textnormal{V}}$ volume is in grains with radius smaller than 0.1 $\mu$m, but 20$\%$ of high $R_{\textnormal{V}}$ volume. For both silicates and carbonaceous grains, as $R_{\textnormal{V}}$ increases, at least 50$\%$ of the volume is in grains of larger and larger size.

The size distributions coming from the posterior resulting from the MCMC are calculated (Fig. \ref{fig:MCMC_size_dist}). They reproduce acceptable size distributions as proposed by WD01 (Fig. \ref{fig:size_dist}).

There is a broad range of parameters that can produce each $R_{\textnormal{V}}$, but the distributions are largely distinct from each other as $R_{\textnormal{V}}$ changes (Fig. \ref{fig:param_bined_distributions}). Each parameter has a different impact on the $R_{\textnormal{V}}-\beta$ anticorrelation (Fig.\ref{fig:param_Rv_vs_beta}). Further work may be warranted to isolate the effect of each parameter on $R_{\textnormal{V}}$ independently of the others.

\begin{figure*}[t]
	\includegraphics[scale=1]{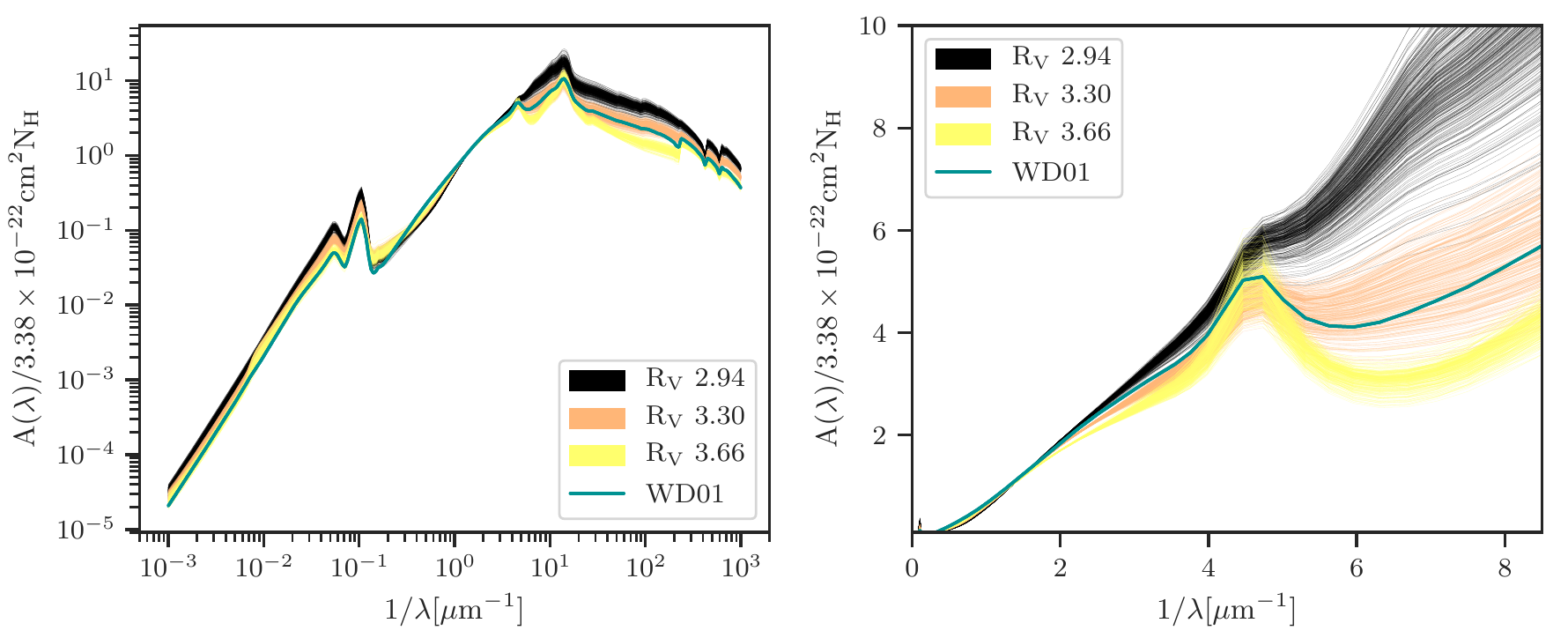}
	\caption{The left panel shows the extinction functions obtained from the MCMC for 3 values of $R_V$ (2.993, 3.300,3.664) for wavelengths ranging from far infrared to x-rays. Looking over the entire wavelength range, we can see that the extinction functions resulting from the MCMC are in agreement with the reference function from WD01. The right panel is a zoomed-in view on the UV feature at  2175 \AA{}; the feature varies between the Milky Way, Small Magellanic Galaxy, and Large Magellanic Galaxy \cite{Fitzpatrick2007}, with the feature being prominent in the Milky Way and less prominent outside of it. The 2175\AA{} bump is often associated with PAHs (\cite{Joblin1992},
		\cite{Li2001}, \cite{Mishra2015}) and in the WD01 model its amplitude is explicitly controlled by the $b_{\textnormal{C}}$ parameter. Since in our study we were aiming to replicate the conditions in the Milky Way, we restricted the $b_{\textnormal{C}}$ parameter to make sure we have a minimum of PAHs involved. }
	\label{fig:MCMC_extinction}
\end{figure*}

\paragraph{Ultraviolet Extinction}\label{sec:UV_extinction}
The model is constrained to match the S16 extinction curve in 10 bands, but is not sufficient to constrain the 2175\AA{} feature.  We compare extinction functions derived from the MCMC with reference functions from WD01 (calculated for the fourth row of Table 1 of WD01) and find good agreement across a wide wavelength range (Fig. \ref{fig:MCMC_extinction}).  In particular the 2175\AA{} feature varies between the Milky Way, Small Magellanic Galaxy, and Large Magellanic Galaxy \cite{Fitzpatrick2007}, with the feature being prominent in the Milky way and less prominent outside of it. The 2175\AA{} bump is often associated with PAHs (\cite{Joblin1992},
\cite{Li2001}, \cite{Mishra2015}) and in the WD01 model its amplitude is explicitly controlled by the $b_{\textnormal{C}}$ parameter. Since in our study we were aiming to replicate the conditions in the Milky Way, we restricted $b_{\textnormal{C}}$ to values greater than 3$\times 10^{-5}$ to make sure we have a minimum amount of PAH involved. The maximum value of $b_{\textnormal{C}}$ is set between (5.0, 6.5) as  $R_{\textnormal{V}}$ goes from low to high since the total volume of the carbonaceous grains increases with $R_{\textnormal{V}}$ as well. The formula used is $\max{b_{\textnormal{C}}}=6\times (1+0.28\times\tanh(\alpha_{V_g}-1))$.

\begin{figure}[t]
	\includegraphics[scale=1]{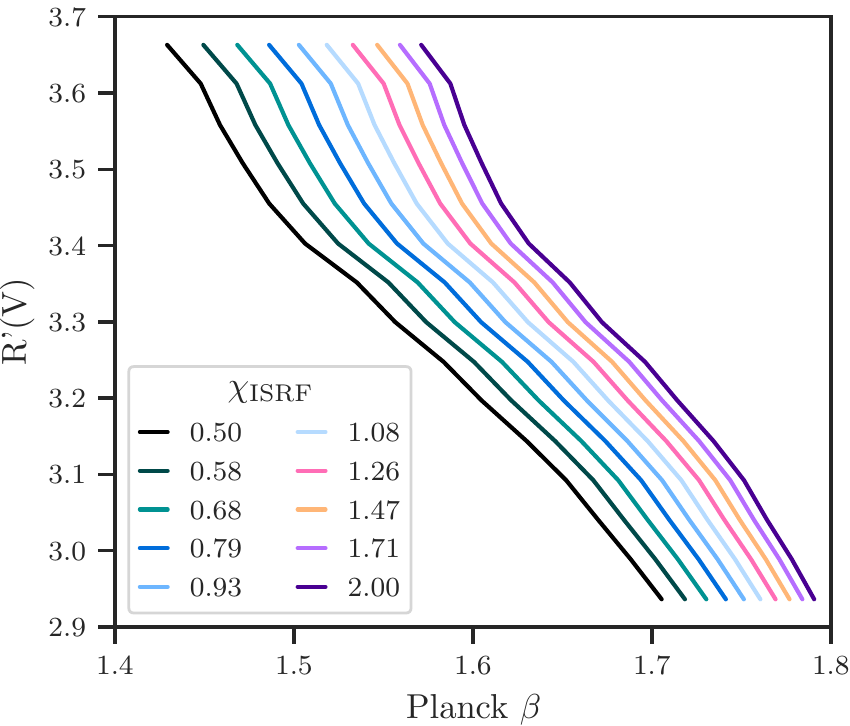}
	\caption{The emissivity of the collection of dust particles for each run is calculated using a modified ISRF multiplication factor $\chi_{\textrm{ISRF}}$.  $\chi_{\textrm{ISRF}}$ takes 10 log spaced values between 0.5 and 2. This results in the $R_{\textnormal{V}}$-$\beta$ correlation being shifted left and right relatively uniformly across $R_{\textnormal{V}}$, with a change of up to 0.1 in $\beta$ at low $R_{\textnormal{V}}$ and 0.15 at high $R_{\textnormal{V}}$. For each  $\chi_{\textrm{ISRF}}$ value, the lines on the plot were generated from the 15 average $R_{\textnormal{V}}$ and $\beta$ value from the MCMC posteriors.}
	\label{fig:Rv_vs_beta_chi}
\end{figure}

\begin{figure}[t]
	\includegraphics[scale=1]{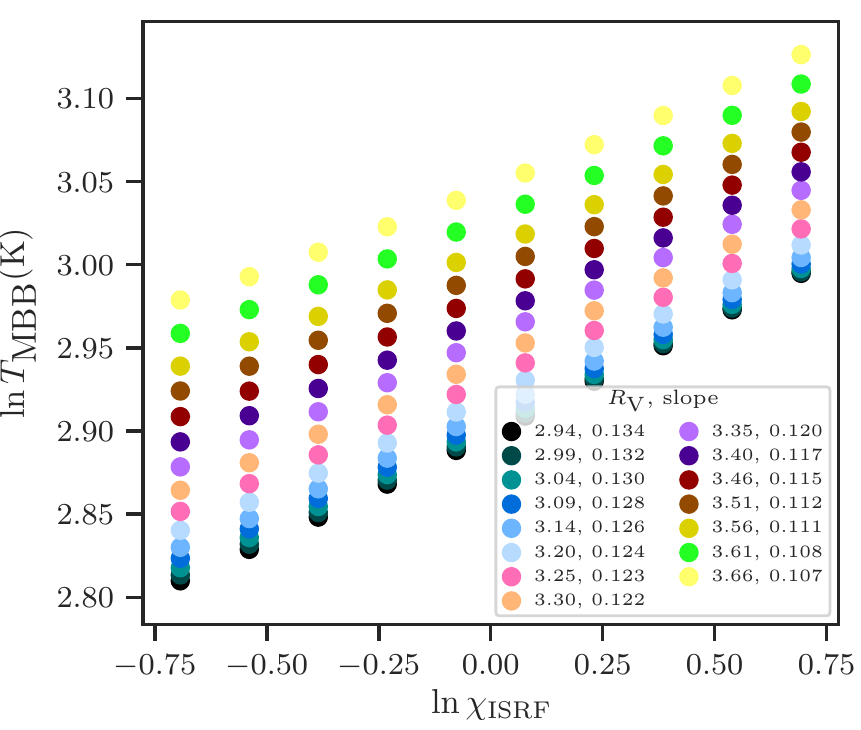}
	\caption{ The logarithm of the average $T_{\textnormal{MBB}}$, for each of the 10 values of $\chi_{\textrm{ISRF}}$, for each of the 15 MCMC runs. This results in a linear relationship between $\ln{T_{\textnormal{MBB}}}$ and $\ln{\chi_{\textrm{ISRF}}}$. The legend shows the slope of the linear fit for each $R_{\textnormal{V}}$ value. The slope decreases with $R_{\textnormal{V}}$, a result due to the variable dust composition, as explained in Section \S \ref{sec:ISRF_effect}.}
	\label{fig:ln_T_vs_ln_chi}
\end{figure}

\begin{figure*}[t!]
	\includegraphics[scale=1]{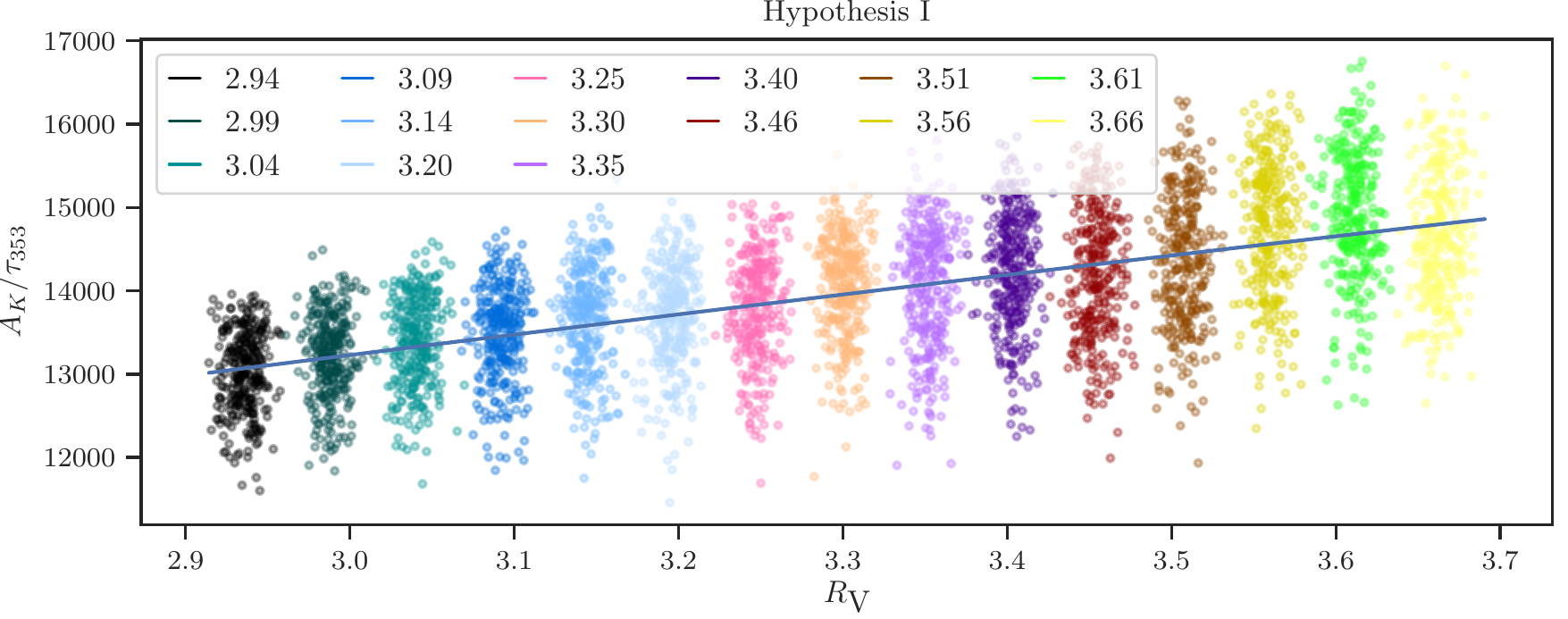}
	
	\includegraphics[scale=1]{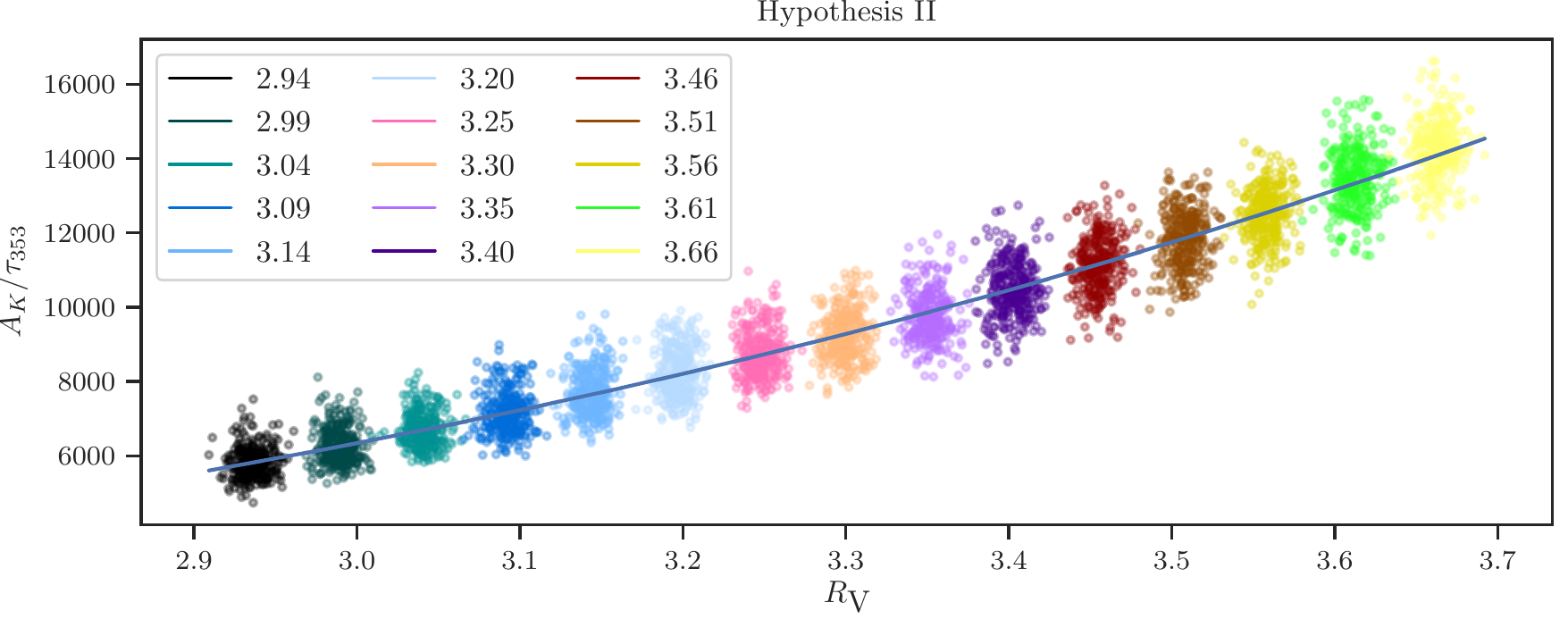}
	\caption{ $R_{\textnormal{V}}$ variation can lead to a significant change in A($\lambda$)/$\tau_{353}$. For the K band shown here, for hypothesis I the power law fit is $A_K/\tau_{353} \propto R_{\textnormal{V}}^{0.58}$, and for hypothesis II it is $A_K/\tau_{353} \propto R_{\textnormal{V}}^{4.00}$.  In this work $R_{\textnormal{V}}$ refers to the $R_{\textnormal{V}}'$ from S16. }
	\label{fig:A_lambda_over_tau_vs_Rv}
\end{figure*}

\subsection{Effect of the Interstellar Radiation Field}\label{sec:ISRF_effect}
The effect of the interstellar radiation field on our analysis is very significant. As described in Section \S \ref{sec:equilibrium_temp}, we assume the ISRF is isotropic and homogeneous, an assumption which is obviously not reflected at the large scales of the universe where variations in proximity, types, and density of stars (and other objects), as well as the cloud thickness, influence the local ISRF dust grains are exposed to. To try to get a glimpse of what the effects of changing the ISRF look like, we modified the ISRF multiplication factor $\chi_{\textrm{ISRF}}$ over a log-spaced array with values between 0.5 and 2 (Figs. \ref{fig:Rv_vs_beta_chi} and \ref{fig:ln_T_vs_ln_chi}). However, the effect is not that drastic, as a factor of 4 in $\chi_{\textrm{ISRF}}$ leads to a change of up to 0.1 in $\beta$ at low $R_{\textnormal{V}}$ and 0.15 at high $R_{\textnormal{V}}$.

In addition, the effect of the variation of the ISRF on the modified black body temperature fit is explored, averaged for each of the 15 MCMC posterior points (Fig. \ref{fig:ln_T_vs_ln_chi}). For each $R_{\textnormal{V}}$ value, a linear relationship is obtained between $\ln{T_{\textnormal{MBB}}}$ and $\ln{\chi_{\textrm{ISRF}}}$. The slope of this linear function decreases as $R_{\textnormal{V}}$ increases. One could expect the slope to be close to $1/(4+\beta)$, and due to the  $R_{\textnormal{V}}$-$\beta$ anticorrelation relation discussed in this paper, the average value of $\beta$ decreases as $R_{\textnormal{V}}$ increases. This would result in the slope increasing at higher $R_{\textnormal{V}}$. However, the discrepancy is explained by the fact that we are doing a modified black body fit over a collection of grains with variable dust composition. For each grain individually (Figs. \ref{fig:power_index} and \ref{fig:gamma_temperature}), the relationship $1/(4+\theta)$ is recovered when calculating the equilibrium temperatures at different $\chi_{\textnormal{ISRF}}$ values. However, the value of the optical power law index $\theta$ varies with the type and size of grain (Fig. \ref{fig:power_index}). At high $R_{\textnormal{V}}$, our runs have a higher ratio of carbonaceous to silicate than at low $R_{\textnormal{V}}$. Since carbonaceous grains have higher $\theta$ than silicate grains, it results in a lower slope obtained when integrating over the contributions from all the particles in the distributions to obtain the modified black body fit.

\begin{figure*}[t!]
	\includegraphics[scale=1]{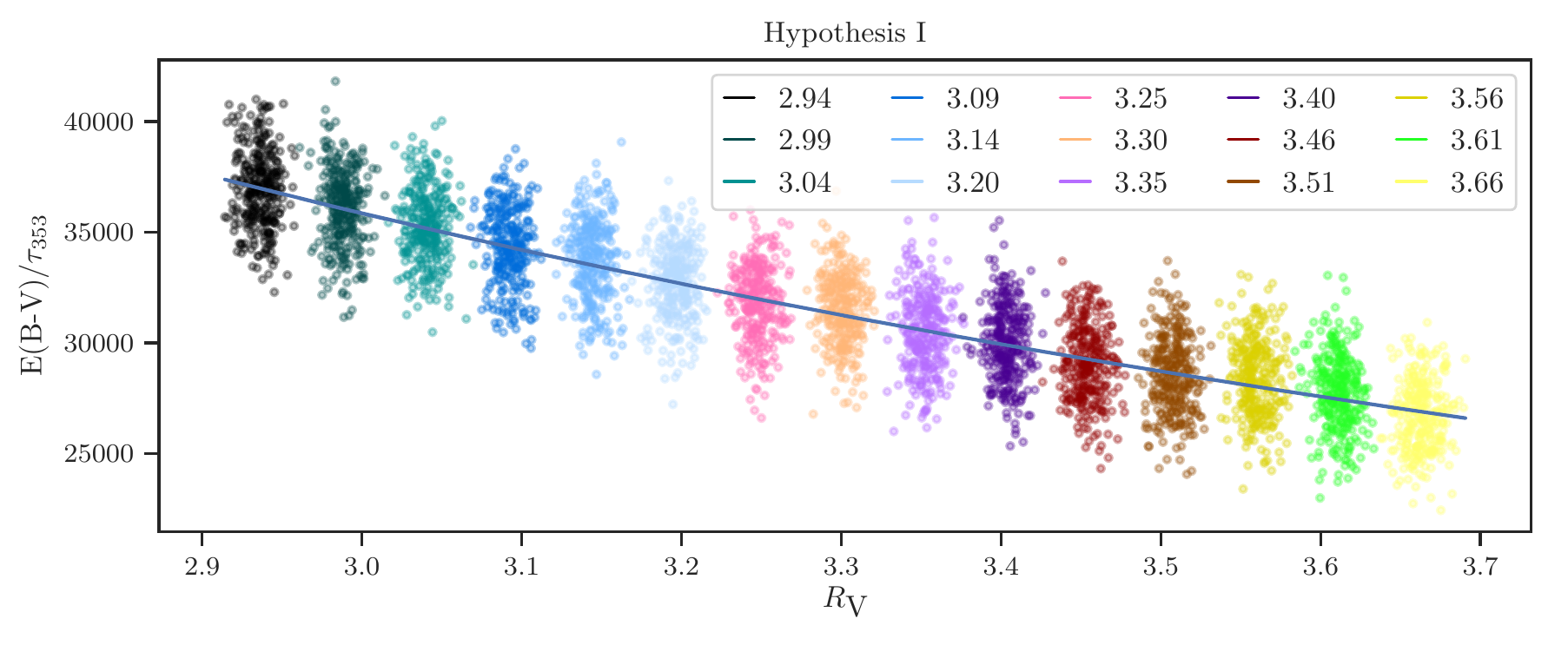}
	\includegraphics[scale=1]{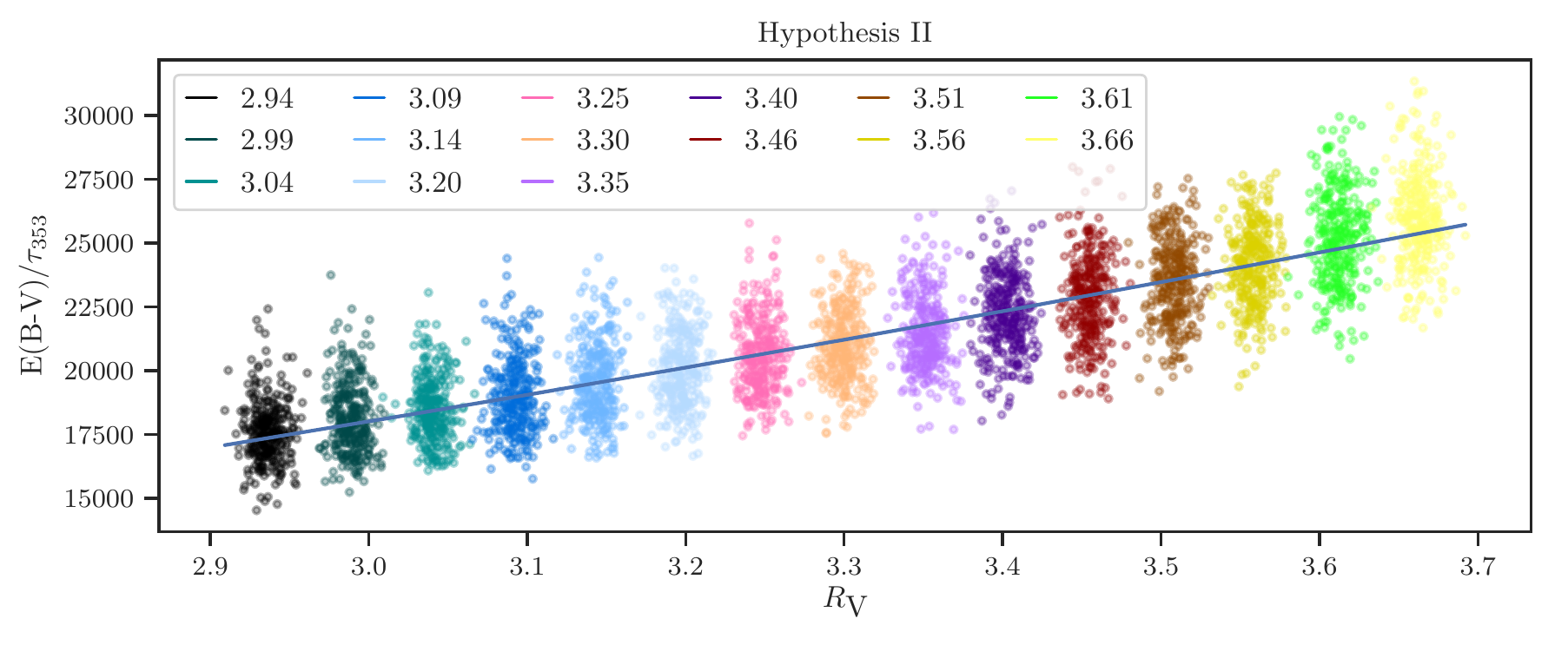}
	\caption{For hypothesis II, a power law fit gives us a variation of the form $E(\textnormal{B}-\textnormal{V})/\tau_{353} \propto R_{\textnormal{V}}^{1.72}$. The 1.72 power law results in values approximately $20\%$ lower at 2.9 and $20\%$ higher at 3.7, compared to $R_{\textnormal{V}}$=3.3. This can have implications for calibrations of emission-based interstellar dust maps. In comparison, for hypothesis I, the power law is $E(\textnormal{B}-\textnormal{V})/\tau_{353} \propto R_{\textnormal{V}}^{-1.44}$. The changing sign of the power index between the case when we keep the volume fixed or varied strengthens the argument for further research. }
	\label{fig:E_B_V_over_tau_vs_Rv}
\end{figure*}

\subsection{Effect on A($\lambda$)/$\tau_{353}$}\label{sec:A_over_tau}

Emission-based interstellar dust maps such as that in \cite{Schlegel1998} have been a very valuable tool for predicting extinction across the sky.
They make the assumption that the ratio of near-infrared extinction to the emission optical depth does not vary with $R_{\textnormal{V}}$.
Using the results for our 15 MCMC runs, we calculate to ratio of A($\lambda$)/$\tau_{353}$ to see how it varies with $R_{\textnormal{V}}$ (Figs. \ref{fig:A_lambda_over_tau_vs_Rv} and \ref{fig:E_B_V_over_tau_vs_Rv}).

We find that there can be a lot of variation in  A($\lambda$)/$\tau_{353}$. For the K band (Fig. \ref{fig:A_lambda_over_tau_vs_Rv}), the best fit power law for hypothesis II is given by

\begin{equation}
\ln{ (A_{\textnormal{K}}/\tau_{353})} = 4.00 \ln{R_{\textnormal{V}}}+4.36.
\end{equation}

For $E(\textnormal{B}-\textnormal{V})/\tau_{353}$ (Fig. \ref{fig:E_B_V_over_tau_vs_Rv}) the power law index is smaller, given by

\begin{equation}
\ln{ (E(\textnormal{B}-\textnormal{V})/\tau_{353})} = 1.72 \ln{R_{\textnormal{V}}}+7.91,
\end{equation}
and there is a tendency for it to be $20\%$ lower at 2.9 and $20\%$ higher at 3.7, compared to $R_{\textnormal{V}}$=3.3. 

We took the $E(\textnormal{B}-\textnormal{V})$ of the stars studied by S16, and the corresponding $\tau_{353}$ data from \cite{Collaboration2016a}, and obtained the average value of $\mu_{E(\textnormal{B}-\textnormal{V})/\tau_{353}}=10,501$. Figure \ref{fig:E_B_V_over_tau_vs_Rv} shows that both hypothesis are above this value. Hypothesis II is closer and thus preferred, but this indicates that the overall dust model could be improved.

This is a potentially impactful result that can motivate future research into the effect of $R_{\textnormal{V}}$ variation on emission-based interstellar dust map calibrations.
The fact that sign of the trend changes depending on whether we assume $R_{\textnormal{V}}$ variation is caused only by size variation, or by size \emph{and} composition variation, suggests that additional research will be required before we can confidently derive extinction from emission-based dust maps as $R_{\textnormal{V}}$ varies.


\begin{figure*}[t!]
	\includegraphics[scale=1]{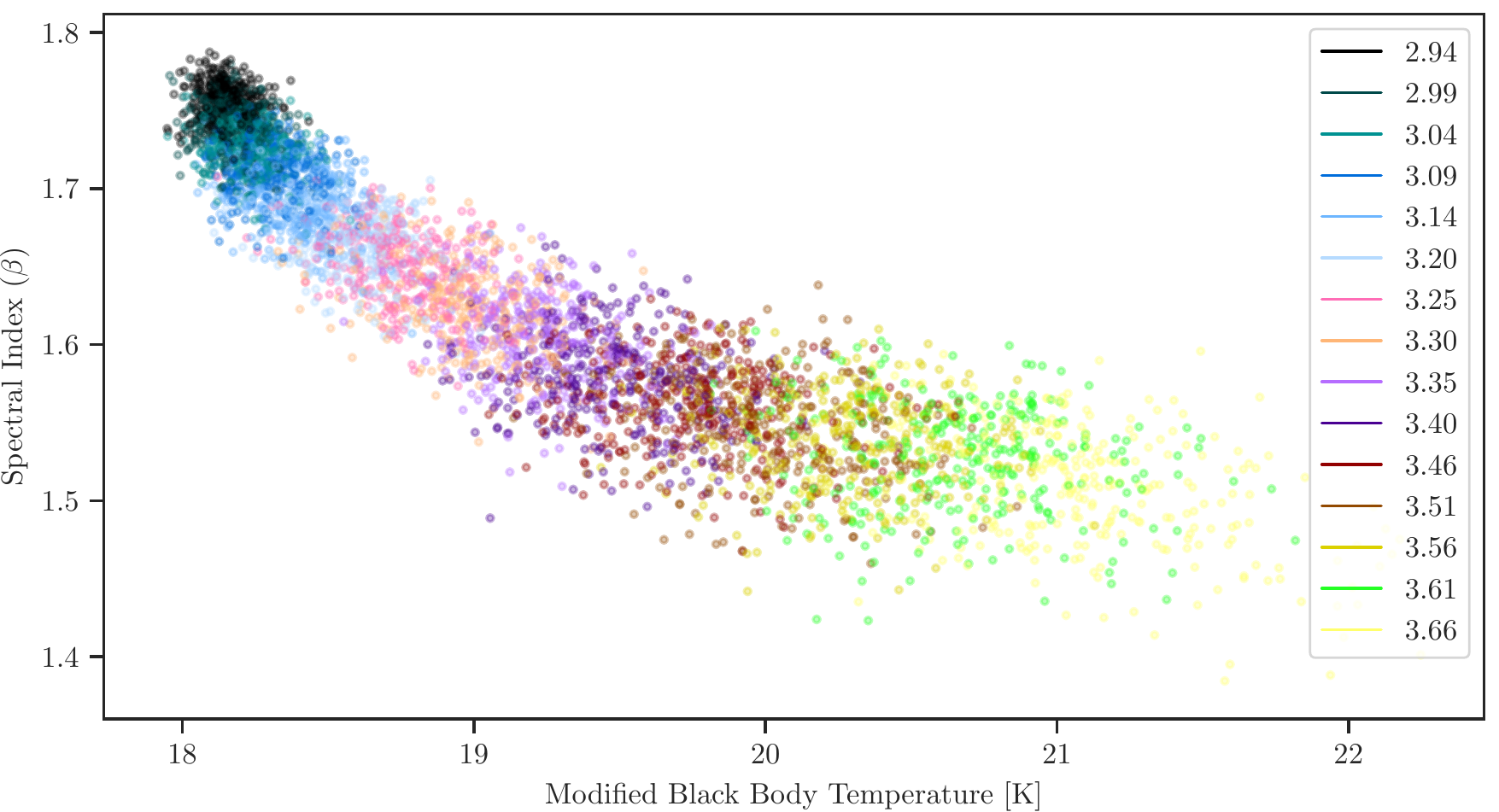}
	\caption{The intensity emission of the collection of dust particles is fit using a modified black body function, for each point in the posterior resulting from the MCMC. Anticorrelation between $\beta$ and $T$ is observed from the resulting distribution of the spectral index ($\beta$) as a function of the collective modified black body teperature ($T$). The values of the temperature are spread between 18-22K. Greater $R_\textnormal{V}$ leads to higher temperature. This is in agreement to the fact that higher $R_\textnormal{V}$ runs have a higher ratio of carbonaceous to silicate grains, since carbonaceous grains have higher individual equilibrium temperature compared to silicates (Fig. \ref{fig:equilibrium_temperature}). }
	\label{fig:beta_vs_T}
\end{figure*}

\subsection{Correlation between spectral index and temperature}

Researchers have been looking at the $T$-$\beta$ correlation and there is not quite a consensus on whether it is real \citep{Dupac2003, Desert2008} or a subtle statistical artifact \citep{Shetty2009,Kelly2012}.  To address this question we consider the modified black body fit to models at varying $R_{\textnormal{V}}$ but with fixed radiation field ($\chi_{\textnormal{ISRF}}$ = 1). It is important to mention that we do not have noise in our model, thus, we cannot comment on studies that presented results on that. In addition, the optical properties of the dust grains can change with the temperature of the grains, and this can lead to a change in the spectral index. However, we do not account for this effect in our models, so we cannot probe this.

The intensity emission of the collection of dust particles is fit using a MBB function, as described in Section \ref{sec:MBB_fit}, for each point in the posterior resulting from the MCMC. When plotting the resulting distribution of the spectral index ($\beta$) as a function of the collective modified black body teperature ($T$) we observe an anticorrelation (Fig. \ref{fig:beta_vs_T}).

One explanation for this interesting result is that $T$ does not affect the long wavelength end very much, so if we boost those points (eg., with more cold dust) to get the $R_{\textnormal{V}}$ correct, the fit must lower $\beta$, but then that lower $\beta$ means a higher $T$ to get the peak in roughly the right place. As a result, there is a natural inverse correlation between $T$ and $\beta$ built into the problem. 

At high $R_\textnormal{V}$ one has larger grains, and larger grains tend to have lower equilibrium temperatures (Fig. \ref{fig:equilibrium_temperature}). However, our runs also have different dust composition. Carbonaceous grains have higher temperature than the silicates, and the ratio of carbonaceous grains to silicate increases as $R_\textnormal{V}$ increases. This explains why in our results, higher $R_\textnormal{V}$ means higher temperature for the modified black body fit over the collection of dust grains.

\section{Conclusion}\label{sec:conclusion}

We started from the size distribution of the grains of dust, and we varied the parameters for it using an MCMC while adding the constraints from the reddening vector obtained by S16. We used the MCMC results to generate the dust emission spectrum for each sample from size distribution space. We suspected that the $R_{\textnormal{V}}-\beta$ correlation would arise naturally from the size distribution alone, but in the case of fixed total volume priors for each dust species, variation of $R_{\textnormal{V}}$ does not produce an appreciable correlation. This is an interesting outcome, and to follow up in the search for a full explanation we force the $R_{\textnormal{V}}-\beta$ and see what parameters give it. We find again that larger grains are correlated with high $R_{\textnormal{V}}$, but in addition we find an explicit function of carbonaceous and silicates volume priors as functions of $R_{\textnormal{V}}$ that gives the $R_{\textnormal{V}}-\beta$ correlation and satisfy the constraints WD01 used. The properties of the optical absorption coefficients for carbonaceous and silicates offer an explanation for the results of the analysis; carbonaceous grains have optical properties that lower the $\beta$ for a collection of dust grains, while silicates raise it (Fig. \ref{fig:power_index}). 

Widely used dust maps like SFD \citep{Schlegel1998} and \cite{Collaboration2016} assume the ratio of $E(\textnormal{B}-\textnormal{V})/\tau_{353}$ is constant. In Section \S \ref{sec:A_over_tau}, we find a dependence of $E(\textnormal{B}-\textnormal{V})/\tau_{353}$ and A($\lambda$)/$\tau_{353}$ on $R_{\textnormal{V}}$. This dependence is a testable consequence of our understanding of the $R_{\textnormal{V}}-\beta$ relation in the context of the WD01 models.
Other optical models and size distribution parameterizations are possible, but if this dependence on $R_{\textnormal{V}}$ persists in future models, it would have serious consequences for the recalibration of emission-based dust maps as a function of $R_{\textnormal{V}}$.

Moreover, this result might provide some guidance on how to improve these dust models in the future. $E(\textnormal{B}-\textnormal{V})/\tau_{353}$ can become an additional constraint used during modeling.  Reproducing the correct function of $E(\textnormal{B}-\textnormal{V})/\tau_{353}$ versus $R_{\textnormal{V}}$ based on real data could be a good target for the next studies.

Modeling the size distribution and composition of dust is an area of active research. The parameterizations of the size distributions and the optical parameters of the grains can be revisited. An alternative model for the size distribution and optical parameters has been proposed by \cite{Zubko2004}, which can be explored in a future work. In the future, we might have to explore grains that are a combination of both carbon and silicate. The model we are using here, though it reproduces many empirical facts about dust, is necessarily a simplification of nature.  Future work may involve other materials, complex grain geometries, composites, and coatings, etc.  Our work is intended as a plausibility argument, not a final determination of parameters for a truly complete model. One might be able to find other solutions that explain the  $R_{\textnormal{V}}-\beta$ correlation. The robust effect we observe is that a composition with higher ratio of carbonaceous to silicate grains leads to more $R_{\textnormal{V}}$ and lower $\beta$. It is an open question whether this tendency is a generic property of all dust models or if it is a specific feature of the precise dust models we are using.

The fact that larger $R_{\textnormal{V}}$ corresponds to smaller silicate volume can be difficult to understand. Denser regions that have larger $R_{\textnormal{V}}$ are expected to have depleted Si, Mg, and Fe from the gas phase. However, in the dense clouds it may not be possible to know how much hydrogen there is. Since we perform our calculations per $N_{\textnormal{H}}$, this could play a significant role. Also, if the carbon is coming out of the gas faster than the silicate is, there might be more carbon per $N_{\textnormal{H}}$ in the dust cloud. Carbonaceous grains could also be misidentified with grains coated with carbons. The exchange of carbon between the solid and gaseous phases of the ISM is not fully understood, but upcoming missions such as SPHEREx \citep{Dore2014} will shed light on this issue.

This work depends critically on the S16 reddening vectors. Their pre-Gaia \citep{Collaboration2016b} analysis of the reddening law was performed in absence of information regarding distances to the stars whose extinction they were modeling. As a result, the absolute extinction cannot be determined, only the relative difference of extinction between bands, after the gray component had been removed from the analysis. This can be improved in future work when the gray component to the extinction can be fixed using Gaia measurements. In addition, this analysis is performed is at fixed $A(\textnormal{I})/N(\textnormal{H})$ , so we still have a free parameter left.  If future data constrains $A(\textnormal{I})/N(\textnormal{H})$ as a function of  $R_{\textnormal{V}}-\beta$, we can modify the volume relations and get the similar results again with different functions, as $A(\textnormal{I})/N(\textnormal{H})$  scale linearly with the  $b_c,C_s,C_g$ parameters combined.

The results of this study provide a possible explanation of the observed $R_{\textnormal{V}}-\beta$ correlation in the context of the WD01, \cite{Laor1993}, \cite{Draine1984}, \cite{Li2001} family of models.  Although this explanation may not be unique, it increases our confidence that the $R_{\textnormal{V}}-\beta$ correlation can be used to our advantage.
For example, the relation can be use as a cross check for CMB experiments: one can start from a sensitive map of the sky in $R_{\textnormal{V}}$, like one created from the datasets from LSST \citep{Collaboration2009}, and determine the corresponding $\beta$. Conversely, one can make predictions of $R_{\textnormal{V}}$ given precise measurements in $\beta$. 
The $R_{\textnormal{V}}-\beta$ correlation provides valuable information about the size distribution and composition of interstellar dust grains, and may lead us toward a more complete model of the interstellar medium. 

\paragraph{Acknowledgments}
We acknowledge helpful conversations with Ana Bonaca, Blakesley Burkhart, Tansu Daylan, Bruce Draine, Cora Dvorkin,  Daniel Eisenstein, John Kovac, Albert Lee, Karin \"{O}berg, Stephen Portillo, Eddie Schlafly, Zachary Slepian, Josh Speagle, Jun Yin, and Catherine Zucker. I.Z. is supported by the Harvard College Observatory. D.F. is partially supported by  NSF grant AST-1614941, “Exploring the Galaxy: 3-Dimensional Structure and Stellar Streams.” 
This research made use of the NASA Astrophysics Data System Bibliographic Services (ADS), the color blindness palette by Martin Krzywinski \& Jonathan Corum\footnote{\url{http://mkweb.bcgsc.ca/biovis2012/color-blindness-palette.png}}, and the Color Vision Deficiency PDF Viewer by Marrie Chatfield \footnote{\url{https://mariechatfield.com/simple-pdf-viewer/}}.

\facilities{Odyssey Cluster, Harvard University}
\software{ptemcee \citep{Vousden2016}, emcee \citep{Foreman-Mackey2013}, NumPy \citep{VanderWalt2011}, Matplotlib \citep{Hunter2007}, pandas \cite{mckinney-proc-scipy-2010}, scikit-learn \citep{Pedregosa2012}, IPython \citep{Perez2007}, Python \citep{Millman2011, Oliphant2007}}

\appendix

\onecolumngrid

\section{Error in Extinction}\label{sec:appendix_error_in_extinction}
We calculate the errors in the extinction function to be used as a reference in the MCMC (Eq. \ref{eq:reference_extinction}).
Denote with $\bm{\epsilon}_0$ and $\bm{\epsilon}_1$ the error vectors in $\bm{R}_0$ and $\dv{\bm{R}}{x}$, respectively. The values for these error vectors are given in Table 2 of S16.

Firstly, we calculate the error propagation in the extinction formula, given by $A(\lambda) = R_0(\lambda) + x \dv{R(\lambda)}{x}$.

Let us assume we have a function $y$ expressed as a linear combination of the variables $x_l$:
\begin{equation}
y(\bm{x}) = \sum_{l}a_lx_l
\end{equation}
Let $\Sigma^{\bm{x}}$ be the covariance matrix for the parameters $x_l$, such that $\Sigma^{\bm{x}}_{kl} = E[(x_l-\mu_{x_l})(x_k-\mu_{x_k})]$. The mean (first moment) of $y$ is then given by equation \ref{eq:first_moment}:
\begin{equation}
\label{eq:first_moment}
E[y] = \mu_y = E\left[\sum_{l}a_lx_l\right] = \sum_la_lE[x_l] = \sum_la_l \mu_{x_l},
\end{equation}
and the variance (second moment) by equation \ref{eq:second_moment}:
\begin{equation}
\label{eq:second_moment}
\begin{split}
\sigma_y^2 &= E \left[(y-\mu_y)^2\right] \\
& = E\left[\left(\sum_la_lx_l - \sum_la_l\mu_{x_l}\right)^2 \right]\\
& = E\left[\left(\sum_la_l(x_l-\mu_{x_l})\right)^2 \right] \\
& =  \sum_{k}^n\sum_{l}^na_ka_l E\left[(x_k-\mu_{x_k})(x_l-\mu_{x_l}) \right] \\
& = \sum_{k}^n\sum_{l}^na_ka_l\Sigma_{kl}^{\bm{x}}
\end{split}
\end{equation}
Let $\Sigma^{\bm{A}}$ be the covariance matrix of $A(\lambda)$. For our case, the coefficient vector is $(1, x)$, and the variable vector $\bm{x} = \left(R_0(\lambda), \dv{R(\lambda)}{x}\right) $. S16 make the assumption that there is no covariance between the errors in the vectors $\bm{R}_0$ and $\dv{\bm{R}}{x}$. As a result, only the diagonal terms of the covariance matrix are nonzero. Then the error $\sigma_A(\lambda)$ is given by equation \ref{eq:sigma_A}: 
\begin{equation}
\label{eq:sigma_A}
\sigma_{A(\lambda)}^2 = \epsilon_0^2(\lambda) + x^2 \epsilon_1^2(\lambda).
\end{equation}

S16 also assume there is no covariance between the errors at different wavelength. As a result, the covariance matrix $\Sigma^{\bm{A}}$ is given by equation \ref{eq:Sigma_A}:
\begin{equation}
\label{eq:Sigma_A}
\Sigma^{\bm{A}}_{kl} = \delta_{kl}\sigma^2_{A(l)}
\end{equation}

Second, we calculate the error introduced by fixing the gray component  $A(\lambda) \rightarrow A(\lambda)+C$ 
such that $A'(H)/A'(K)=1.55= r$ \citep{Indebetouw2005}:
\begin{equation}
\begin{split}
&A'(\lambda) = A(\lambda)+C\\
&C = \frac{1}{r-1}A(H)-\frac{r}{r-1}A(K)\\
&A'(\lambda) = A(\lambda) + \frac{1}{r-1}A(H)-\frac{r}{r-1}A(K)
\end{split}
\end{equation}

We calculate the error in $A'(\lambda)$. Since the errors in the different bands from the reddening vector were not correlated, neither are the errors of $A(\lambda)$.
The $r$ parameter comes from a measurement given with $10\%$ precision. We keep $r$ fixed in this analysis. The effect of having $r$ values with $\pm 10\%$ difference can be easily explored by fixing $r$ to different values and rerunning the analysis.

The covariance matrix of $A'(\lambda)$ is given by equation \ref{eq:A'_covariance}:
\begin{equation}
\label{eq:A'_covariance}
\begin{split}
\Sigma^{\bm{A'}}_{kl} &=E[(A'(k)-\mu_{A'(k)})(A'(l)-\mu_{A'(l)})] \\
&=\Sigma^{\bm{A}}_{kl} + \frac{1}{r-1}\left(\Sigma^{\bm{A}}_{kH} + \Sigma^{\bm{A}}_{lH}\right)
-\frac{r}{r-1}\left(\Sigma^{\bm{A}}_{kK}+\Sigma^{\bm{A}}_{lK}\right) 
+\frac{1}{(r-1)^2}\Sigma^{\bm{A}}_{HH}
+\left(\frac{r}{r-1}\right)^2\Sigma^{\bm{A}}_{KK}
\end{split}
\end{equation}

Due to fixing the gray component, the covariance matrix of $A'(\lambda)$ is not diagonal, but it is still symmetrical.

The indexes of the matrix run over the 10 filters, in the order g,r,i,z,y,J,H,K,W1,W2. Labeling with $b=\frac{\sigma_{A(H)}^2+r^2\sigma_{A(K)}^2}{(r-1)^2}$, $c=\frac{r\sigma_{A(H)}^2+r^2\sigma_{A(K)}^2}{(r-1)^2}$, $d=\frac{\sigma_{A(H)}^2+r\sigma_{A(K)}^2}{(r-1)^2}$:
\begin{equation}
\Sigma^{\bm{A'}} = 
\begin{bmatrix}
\sigma_{A(g)}^2 + b & b & b & b & b & b & c&d&b&b\\
b & \!\!\!\!\! \sigma_{A(r)}^2 + b & b & b & b & b & c&d&b&b\\
b & b &\!\!\!\!\! \sigma_{A(i)}^2 + b & b & b & b & c&d&b&b\\
b & b & b &\!\!\!\!\! \sigma_{A(z)}^2 + b & b & b & c&d&b&b\\
b & b & b & b &\!\!\!\!\! \sigma_{A(y)}^2 + b & b & c&d&b&b\\
b & b & b & b & b &\!\!\!\!\! \sigma_{A(J)}^2 + b & c&d&b&b\\
c & c & c & c & c & c &\!\!\!\!\! \frac{r^2\left(\sigma_{A(H)}^2+\sigma_{A(K)}^2\right)}{(r-1)^2}&\!\!\!\frac{r\left(\sigma_{A(H)}^2+\sigma_{A(K)}^2\right)}{(r-1)^2}&c&c\\
d & d & d & d & d & d&\!\!\!\!\! \frac{r\left(\sigma_{A(H)}^2+\sigma_{A(K)}^2\right)}{(r-1)^2}&\!\!\!\frac{\sigma_{A(H)}^2+\sigma_{A(K)}^2}{(r-1)^2}&d&d\\
b & b & b & b & b & b& c&d&\!\!\!\!\!\sigma_{A(W1)}^2 + b&b\\
b & b & b & b & b & b& c&d&b&\!\!\!\!\!\sigma_{A(W2)}^2 + b\\
\end{bmatrix}
\end{equation}

Due to fixing of the grey component by requiring $A(H)/A(K)= r$, the rows (and columns) of H and K in the covariance matrix are now related by the constant $r$, making the covariance matrix singular. As such, we remove the row and column corresponding to the H band, and redefine the matrix in equation \ref{eq:Sigma_A'_removed}:

\begin{equation}
\label{eq:Sigma_A'_removed}
\Sigma^{\bm{A'}} = 
\begin{bmatrix}
\sigma_{A(g)}^2 + b & b & b & b & b & b &d&b&b\\
b & \!\!\!\!\! \sigma_{A(r)}^2 + b & b & b & b & b &d&b&b\\
b & b &\!\!\!\!\! \sigma_{A(i)}^2 + b & b & b & b &d&b&b\\
b & b & b &\!\!\!\!\! \sigma_{A(z)}^2 + b & b & b &d&b&b\\
b & b & b & b &\!\!\!\!\! \sigma_{A(y)}^2 + b & b &d&b&b\\
b & b & b & b & b &\!\!\!\!\! \sigma_{A(J)}^2 + b &d&b&b\\
d & d & d & d & d & d&\!\!\!\frac{\sigma_{A(H)}^2+\sigma_{A(K)}^2}{(r-1)^2}&d&d\\
b & b & b & b & b & b& d &\!\!\!\!\!\sigma_{A(W1)}^2 + b&b\\
b & b & b & b & b & b& d &b&\!\!\!\!\!\sigma_{A(W2)}^2 + b\\
\end{bmatrix}
\end{equation}

The third step is to normalize at the third bandpass value of $7572$ \AA{} $\  =  0.7572\ \mu$m, corresponding to the I filter. This is done in order to fix the extinction per hydrogen column density ($N_{\textnormal{H}}$) to the chosen prior convention value at the $I$ band. As a result, we want to divide by the average of $A'(I)$, $\mu_{A'(I)}$, and multiply by our chosen prior convention value of $C_{\frac{A_I}{N_{\textnormal{H}}}}$: 
 
\begin{equation}
\label{eq:A''}
A''(\lambda)=\frac{A'(\lambda)}{\mu_{A'(I)}}C_{\frac{A_I}{N_{\textnormal{H}}}}
\end{equation}

Since we treat the average of $A'(I)$ as a fixed quantity without errors, the covariance matrix for the elements of the vector  $A''(\lambda)$ is given by Equation \ref{eq:A''_cov}:

\begin{equation}
\label{eq:A''_cov}
\Sigma^{\bm{A''}}_{kl} = \frac{\Sigma^{\bm{A'}}_{kl}}{\mu_{A'(I)}^2}C_{\frac{A_I}{N_{\textnormal{H}}}}^2
\end{equation}

\section{Extending the dust optical properties} \label{sec:extend}

\begin{figure}[t!]
	\includegraphics[scale=1]{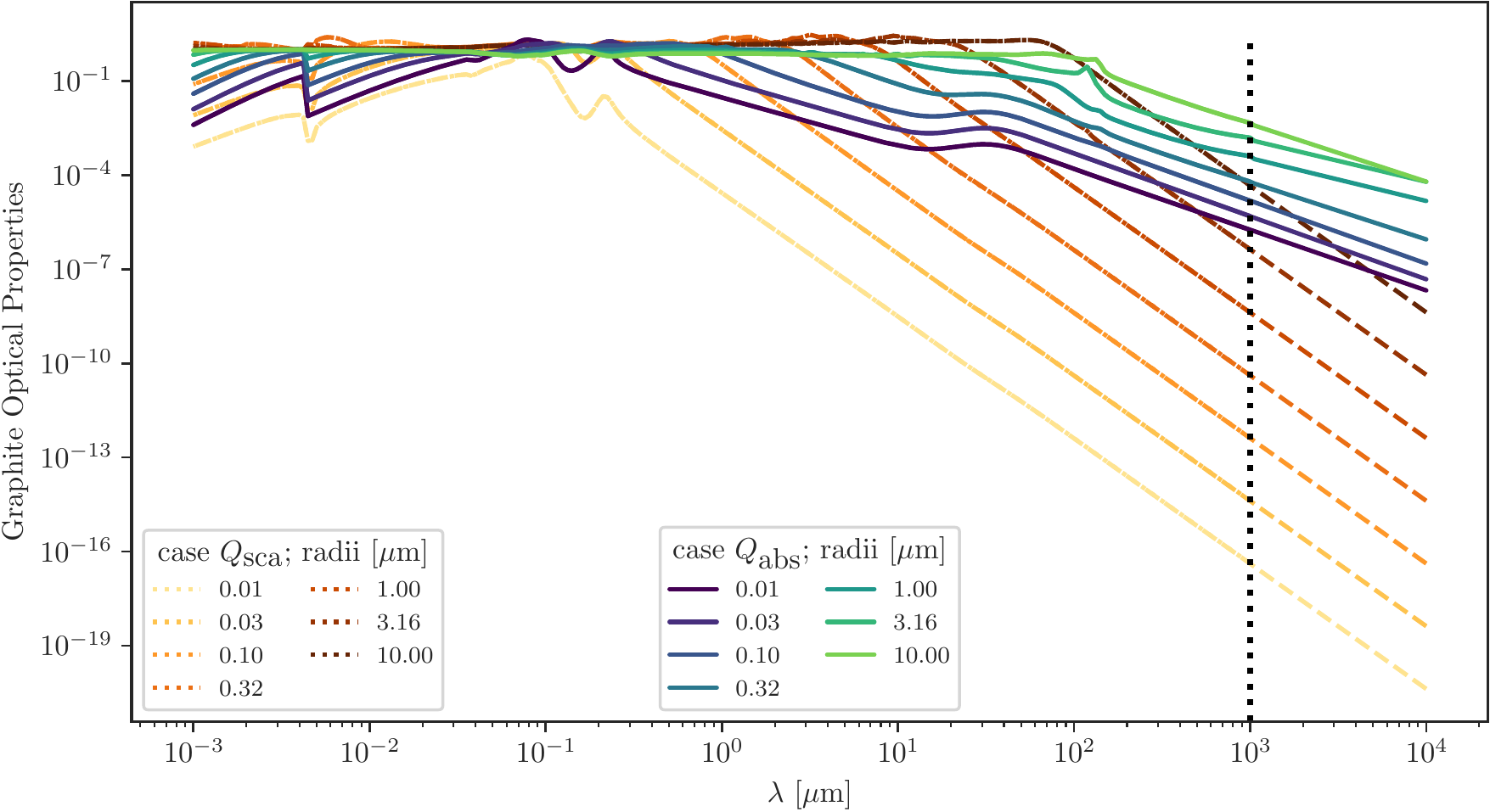}
	\caption{Extending the graphite optical properties for absorption and scattering between $10^3\mu$m and $10^4 \mu$m. The gray-dotted line indicates the boundary of the extension.}
	\label{fig:Q_graphite}
\end{figure}

\begin{figure}[t!]
\includegraphics[scale=1]{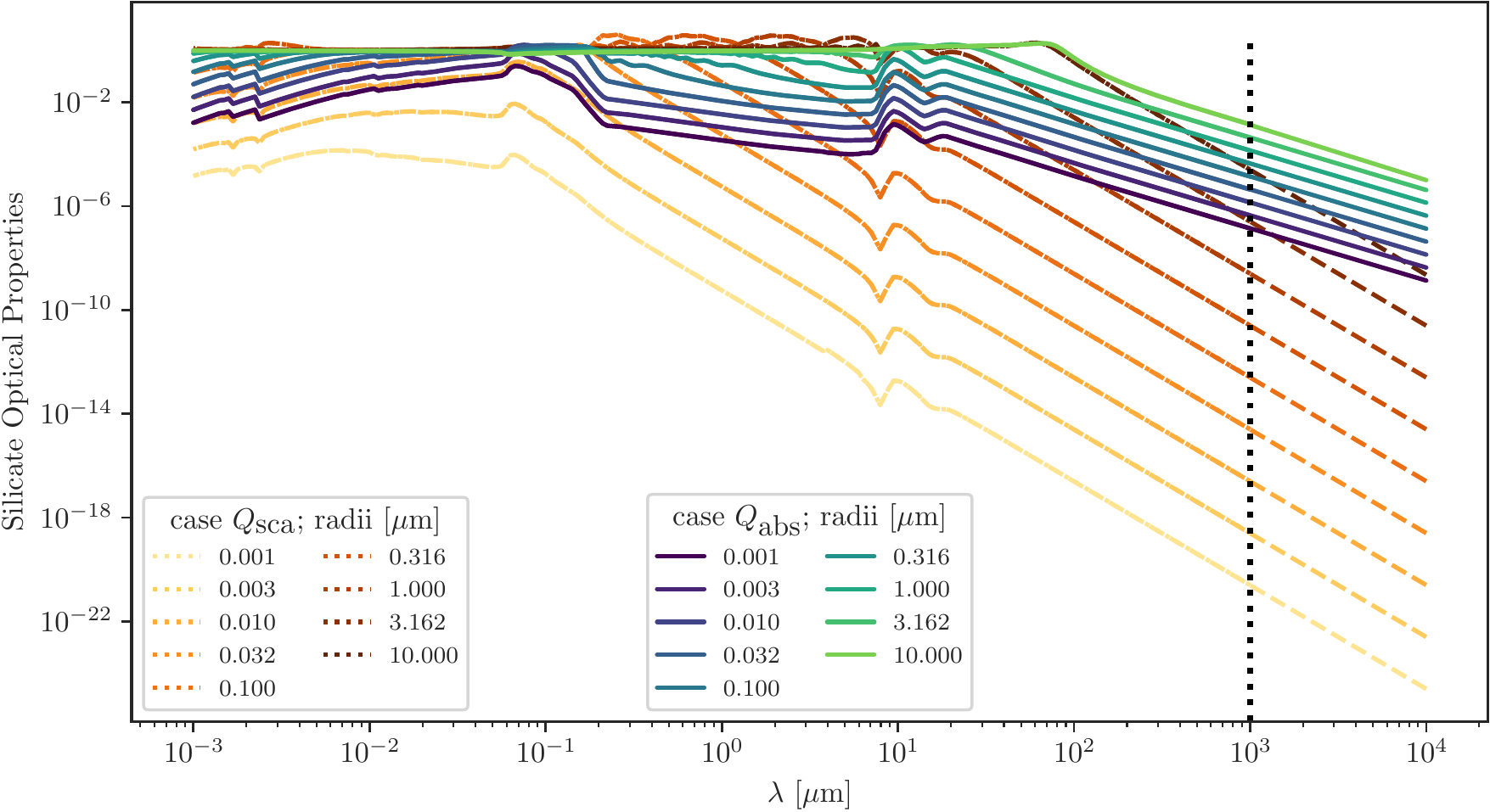}
\caption{Extending the silicate optical properties for absorption and scattering between $10^3\mu$m and $10^4 \mu$m. The gray-dotted line indicates the boundary of the extension. \label{fig:Q_silicate}}
\end{figure}

\begin{figure}[t!]
\includegraphics[scale=1]{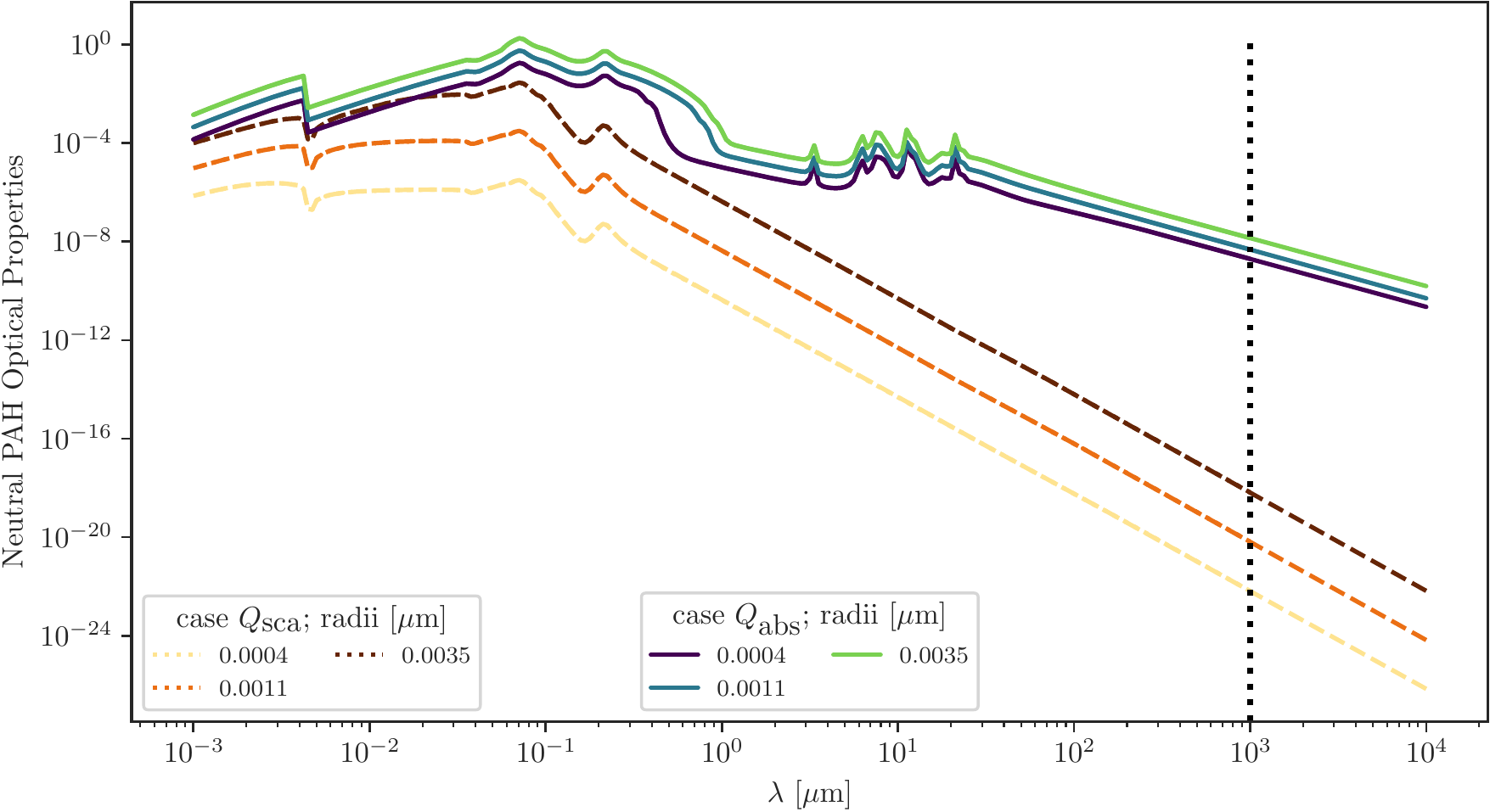}
\caption{Extending the neutral PAH optical properties for absorption and scattering between $10^3\mu$m and $10^4 \mu$m. The gray-dotted line indicates the boundary of the extension. \label{fig:Q_PAH_neu}}
\end{figure}

\begin{figure}[t!]
\includegraphics[scale=1]{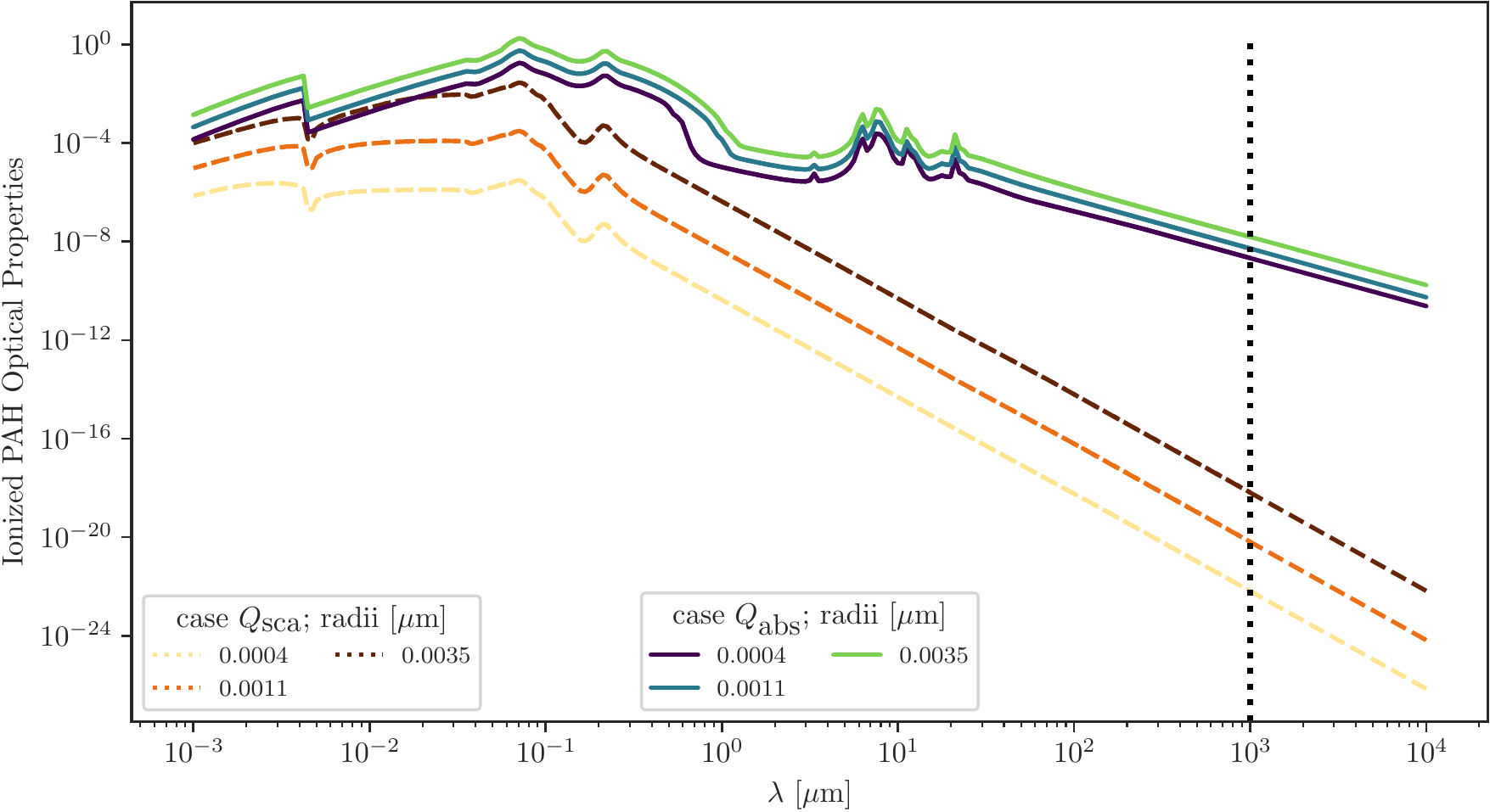}
\caption{Extending the ionized PAH optical properties for absorption and scattering between $10^3\mu$m and $10^4 \mu$m. The gray-dotted line indicates the boundary of the extension.  \label{fig:Q_PAH_ion}}
\end{figure}

The optical properties of the dust grains were extended beyond the wavelengths given by \cite{Laor1993}, \cite{Draine1984}, and \cite{Li2001}, to values between $10^3 \mu$m and $10^4 \mu$m. The extension was done by modeling $Q_{\textnormal{radii}}(\lambda) = \tau ({\lambda/\lambda_0})^{-\theta} $, with $\lambda_0=1$mm and fitting for the $\theta$ and $\tau$ parameters for each radii, using the last 20 bins. Then, the optical parameters were calculated for the new range. The calculation was performed separately for the scattering and absorption coefficients (Figs. \ref{fig:Q_graphite}, \ref{fig:Q_silicate}, \ref{fig:Q_PAH_neu} and \ref{fig:Q_PAH_ion}).

\bibliography{dust.bib}
\end{document}